\documentclass[preprint,english,showpacs,preprintnumbers,amsmath,amssymb,floatfix]{revtex4}
\usepackage[T1]{fontenc}
\usepackage[latin9]{inputenc}
\usepackage{color}
\usepackage{array}
\usepackage{amstext}
\usepackage{graphicx}
\usepackage{esint}

\makeatletter

\@ifundefined{textcolor}{}
{%
 \definecolor{BLACK}{gray}{0}
 \definecolor{WHITE}{gray}{1}
 \definecolor{RED}{rgb}{1,0,0}
 \definecolor{GREEN}{rgb}{0,1,0}
 \definecolor{BLUE}{rgb}{0,0,1}
 \definecolor{CYAN}{cmyk}{1,0,0,0}
 \definecolor{MAGENTA}{cmyk}{0,1,0,0}
 \definecolor{YELLOW}{cmyk}{0,0,1,0}
 }

\@ifundefined{definecolor}
 {\usepackage{color}}{}
\@ifundefined{definecolor}
 {\usepackage{color}}{}
\makeatother

\makeatother

\usepackage{babel}

\begin{document}
%
\newcommand{\MSbar}{\ensuremath{\overline{\text{MS}}\ }}

\title{The pQCD analysis to extract PDFs and $\alpha_s^{{\rm NLO}}(M^2_Z)$ from  	
inclusive jet-hadron production data }

\author{A.~Vafaee}
\email[]{vafaee.phy@gmail.com}
\affiliation{Iran's National Elites Foundation, P. O. Box 14578-93111, Tehran, Iran}
\author{A.~B.~Shokouhi}
\email[]{shokouhi.phy@gmail.com}
\affiliation{Independent researcher, P. O. Box 11155-811, Tehran, Iran}

\date{\today}

\begin{abstract}
This perturbative Quantum Chromo Dynamics (pQCD) analysis attempts to present a simultaneous determination of parton distribution functions (PDFs) and the strong coupling $\alpha_s(M^2_Z)$ from inclusion of inclusive H1 and ZEUS jet, DiJets and TriJets production cross sections data on the HERA I and II combined data, as the central data sets for probing the internal structure of proton. To present an accurate pQCD analysis, we separate the role and influence of inclusion jet production cross sections data from inclusion of the strong coupling $\alpha_s(M^2_Z)$ as an extra fit parameter parameter on the gluon distribution. We show inclusion of jet, DiJets and TriJets production cross sections data improves the consistency between experiment and theory of cross sections for neutral current (NC) and charged current (CC) interactions of deep inelastic ${e^\pm}p$ scattering on proton up to $\sim 2.8$~\%. In addition, we show inclusion of jet production cross section data and considering $\alpha_s(M^2_Z)$ as a pQCD free parameter not only reduce dramatically the uncertainty band of gluon distribution but also improve the consistency between experiment and theory of NC and CC deep inelastic ${e^\pm}p$ scattering cross sections up to $\sim 3.6$~\%. Our simultaneous determination of PDFs and the strong coupling $\alpha_s(M^2_Z)$ with inclusion of jet production cross sections data leads to $\alpha_s^{{\rm NLO}}(M^2_Z) = 0.12041 \pm 0.00086$, which is in a good agreement with world average and other individual measurements.
  
\end{abstract}


\maketitle

\section{\label{introduction}Introduction}
We may qualitatively introduce the potential function between a quark ($q$) and an antiquark ($\bar{q}$) for virtual pair $q\bar{q}$ within the proton as follows:

\begin{eqnarray}
V(r) = -(\frac{4}{3})\frac{\alpha_s}{r} + kr~. \label{eq:jet}
\end{eqnarray}
In qualitative Eq~(\ref{eq:jet}), the first term is a Coulomb potential and for distances much shorter than of a typical hadron size $\sim\frac{1}{m_\pi}$ dominates and the second term is known as the confining potential, which has a nonperturbative origin and probably comes from gluon self-couplings. The second term ($kr$) produces a force that does not reduce with distance and accordingly a quark-antiquark pair ($q\bar{q}$) are tied together by a kind of force-string which effectively requires infinite energy to separate them from each other, a phenomenon which is referred to as the quark confinement~\cite{Vega:2018eiq,Efimov:1988aub,delaRipelle:1989xpy,Blaha:1975fde,Proceedings:2016bba,Wilson:1974sk,Ayala:2015axa,Brambilla:1999ja}.

Really, because of quark confinement phenomenon, quarks are never observed as free particles, but are
always found confined within hadrons. However, in Deep Inelastic Scattering (DIS) of $e^{\pm}p$ processes it is quarks that are produced, not hadrons. Accordingly, when a quark-antiquark pair ($q\bar{q}$) is produced in the centre of frame of a $e^{+}e^{-}$ system through the process like $e^{+}e^{-}\rightarrow q\bar{q}$ they fly apart at relativistic velocities in opposite directions with equal momentum~\cite{Aaron:2009vs,Frixione:1997ks,Andersson:1983ia,Webber:1993bm}.

As a result of quark confinement, the energy in the strong interaction field between the two
quarks is converted into further pairs of $q\bar{q}$ through a process called hadronisation that occurs over a distance scale of $\sim1$~fm. Really, when the distance between $q\bar{q}$ pair exceeds $\sim1$~fm the stored energy in the strong interaction field between them exceeds the mass energy of typical hadrons and the creation of an extra hadron is energetically favored over that of stretching the string further. Many hadrons are created in this manner on the line between the original $q\bar{q}$ pair. They are dragged by their own parent $q$ or $\bar{q}$ and are emitted in concentration in the opposite directions which are observed as jets. Therefore, as a result of hadronisation, each quark produced in a DIS of $e^{\pm}p$ collision produces a jet of hadrons and accordingly a quark is observed as an energetic jet of particles~\cite{Barone:1999yv,Dasgupta:2009tm,Chekanov:2002be}.

The factorisation theorem separates the pQCD short and long distances processes to hard scattering coefficients and PDFs. The hard scattering coefficients and PDFs are calculable and non-calculable parts of pQCD, respectively~\cite{Abramowicz:2015mha,Collaboration:2010ry,Andreev:2013vha,Aaron:2012qi,Adloff:2003uh,Adloff:2000qj,Aaron:2009aa,Aaron:2009bp,Aaron:2009kv,Adloff:1999ah,Gribov:1972ri,Gribov:1972rt,Altarelli:1977zs}.

 Proton structure functions are then obtained by convolution between hard scattering coefficients and PDFs. Since proton PDFs are non-calculable part of pQCD, they are first parametrized at a starting scale of $Q_0^{2}$ based on a standard PDF model and then extracted from fit to experimental data~\cite{Breitweg:1997hz,Breitweg:2000yn,Chekanov:2001qu,Dokshitzer:1977sg,CooperSarkar:1987ds,Derrick:1996hn,Adloff:1997mf,Aaron:2009af,Benvenuti:1989rh,Arneodo:1996qe}. 
 
This next-to-leading order (NLO) QCD analysis attempts to extract simultaneously PDFs and the strong coupling $\alpha_s(M^2_Z)$ from three different data sets: $1$-~HERA I and II combined data~\cite{Abramowicz:2015mha}, $2$-~H1 normalized inclusive jet data~\cite{Aaron:2009vs} and $3$-~ZEUS inclusive jet data sets~\cite{Chekanov:2002be}. The inclusion of H1 and ZEUS inclusive jet-production cross sections made a simultaneous determination of PDFs and the strong coupling $\alpha_s(M^2_Z)$, resulting in the variant HERAPDF2.0Jets.

Jet production in neutral current (NC) of DIS of $e^{\pm}p$ collision at high $Q^2$ provides a testing ground for the theory of the strong interaction between quark and gluon intraction in quantum chromodynamics level~\cite{Aaron:2009vs}.

 While inclusive DIS of $e^{\pm}p$ collision gives us some indirect information on the strong coupling $\alpha_s(M^2_Z)$ through scaling violations of the proton structure functions but jet production provides a direct estimate of $\alpha_s(M^2_Z)$ as an important pQCD parameter. Accordingly, in the most QCD analysis to extract the PDFs, the strong coupling $\alpha_s(M^2_Z)$ is not taken account as a free QCD parameter and uncertainty band of gluon distribution is reduced for fitting with fixed $\alpha_s(M^2_Z)$ compared to fitting with free $\alpha_s(M^2_Z)$. This is our main motivation to perform a NLO DIS of $e^{\pm}p$ collision analysis with inclusion of inclusive jet production data sets and the strong coupling $\alpha_s(M^2_Z)$ as an important pQCD free parameter on the theory of NC and CC deep inelastic ${e^\pm}p$ scattering cross sections to investigate the pure impact of inclusion of the inclusive jet production data on simultaneous determination of $\alpha_s^{{\rm NLO}}(M^2_Z)$ and the gluon distribution.

In perturbative QCD fits to the inclusive HERA I and II combined NC and CC deep $e^{\pm}p$ scattering cross section data as a central data for probing the internal structure of proton  as a whole, the gluon PDF is determined by the DGLAP collinear evolution equations using the observed scaling violations~\cite{DGLAP}. The results of this fitting are in a strong correlation between the shape of the gluon distribution and the numerical values of the strong coupling $\alpha_s(M^2_Z)$, when it is considered as a free fit parameter~\cite{Aktas:2007aa,Dissertori:2007xa,Chekanov:2005ve,Jones:2003yv,Dasgupta:1997ex}.

Inclusive jet production DIS cross section data provide an independent measurement of PDFs and special for gluon distribution and some of its relative ratios and we show inclusion of these data on the theory of cross sections for NC and CC interactions of deep inelastic ${e^\pm}p$ scattering reduces the uncertainty band of gluon distribution and simultaneously provides an accurate determination of the $\alpha_s(M^2_Z)$, when it is considered as a pQCD fit parameter.

To present an accurate study of NLO QCD analysis, we separate the role and influence of inclusion of inclusive jet production DIS cross section data from the role of strong coupling $\alpha_s(M^2_Z)$ on the gluon distribution. To this purpose we develop four different fits with fixed and free $\alpha_s(M^2_Z)$ to investigate the pure impact of inclusion of the inclusive jet production data to the HERA I and II combined NC and CC deep $e^{\pm}p$ scattering cross section data sets as central data for proton structure as a whole and performing an accurate simultaneous determination of $\alpha_s^{{\rm NLO}}(M^2_Z)$ and the gluon distribution. 

 The outline of this  paper is as follows: In Sec.~(\ref{dis}) we describe the theory of NC and CC deep inelastic $e^{\pm}p$ scattering and inclusive jet production cross sections . We introduce our fit methodology and QCD set-up in Sec.~(\ref{fit}). We present our NLO QCD analysis results in Sec.~(\ref{results}) and then we conclude with a summary in Sec.~(\ref{summary}).

\section{\label{dis}Theory and Cross Sections}

\begin{figure*}
\includegraphics[width=0.49\textwidth]{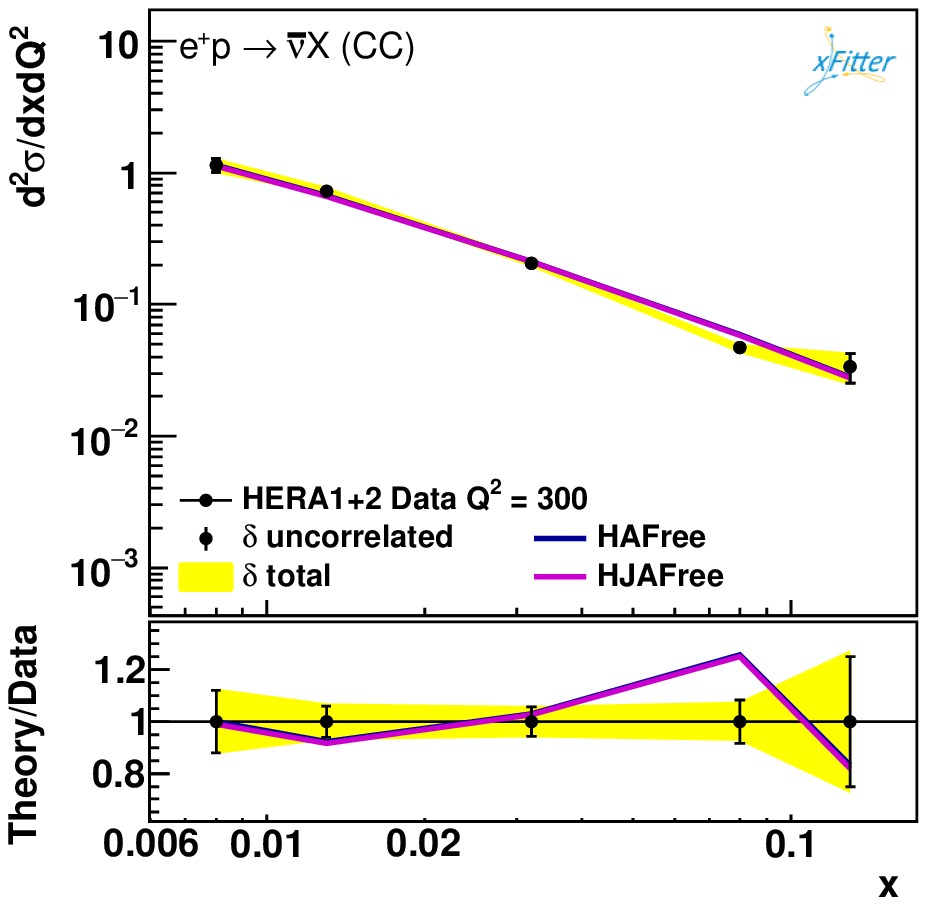}
\includegraphics[width=0.49\textwidth]{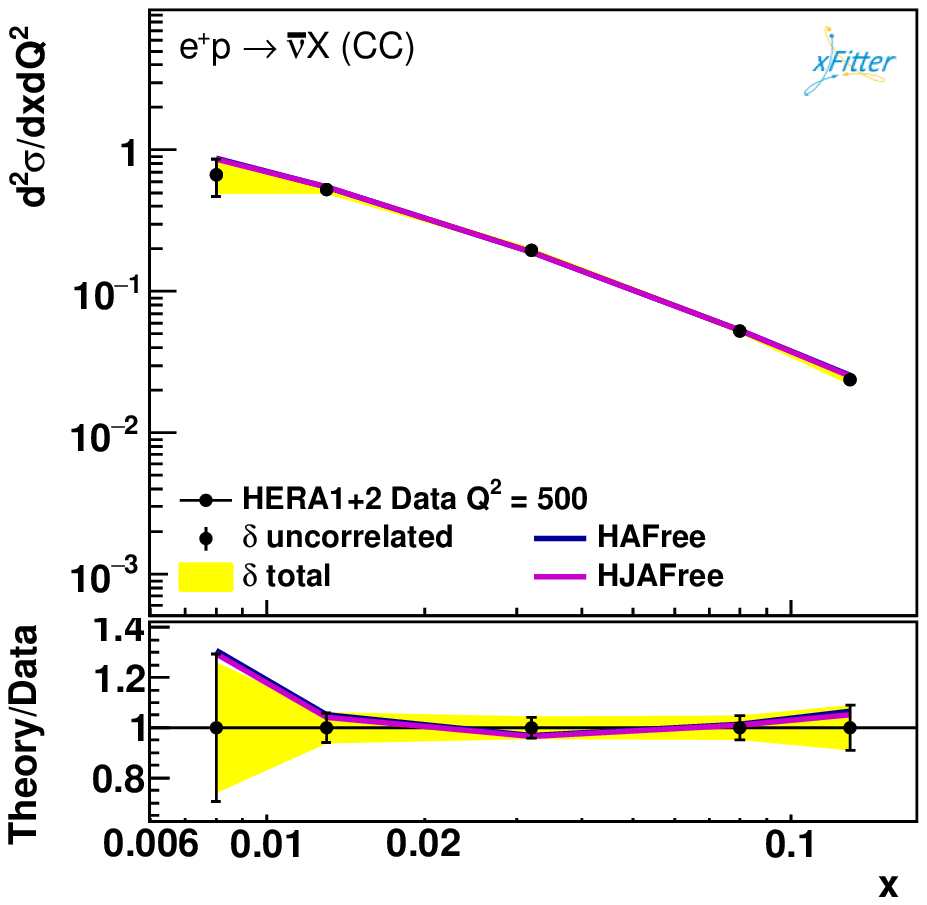}

\caption{The CC deep inelastic ${e^+}p$ scattering double-differential cross sections $\frac{d^2\sigma_{CC}^{e^{\pm} p}}{d{x}dQ^2}$ as a function of $x$ and consistency of data with theory predictions at low $Q^2$.}
\label{fig:1}
\end{figure*}

\begin{figure*}
\includegraphics[width=0.49\textwidth]{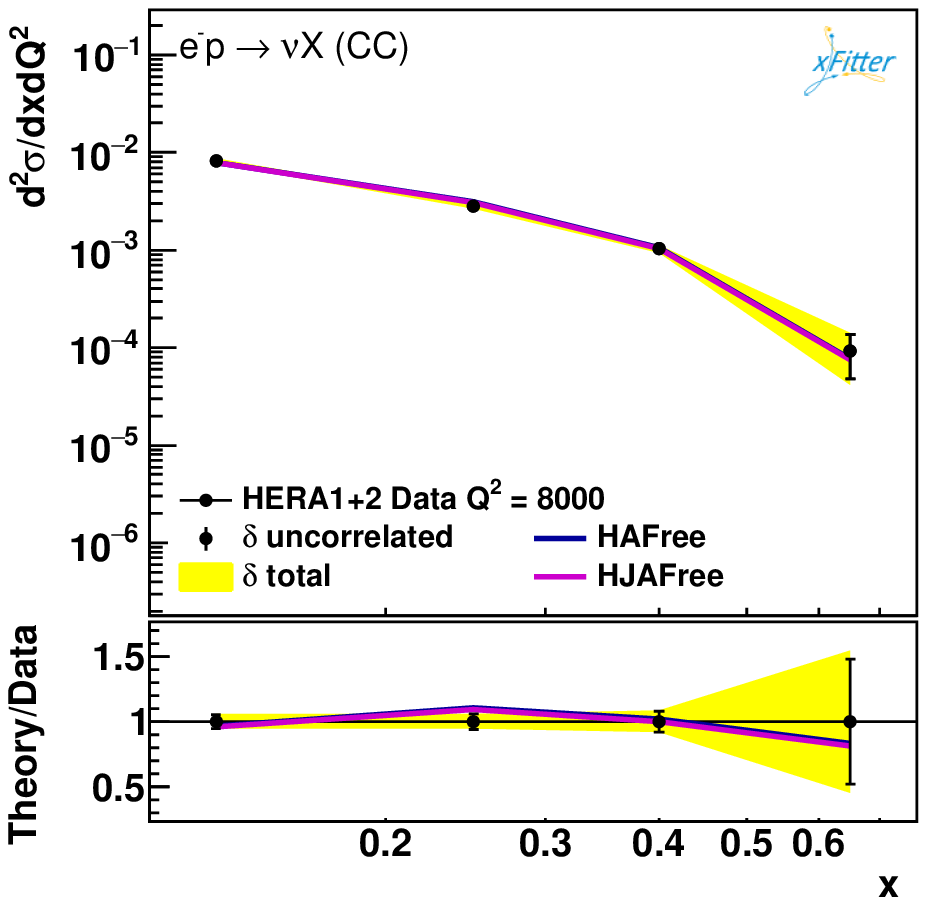}
\includegraphics[width=0.49\textwidth]{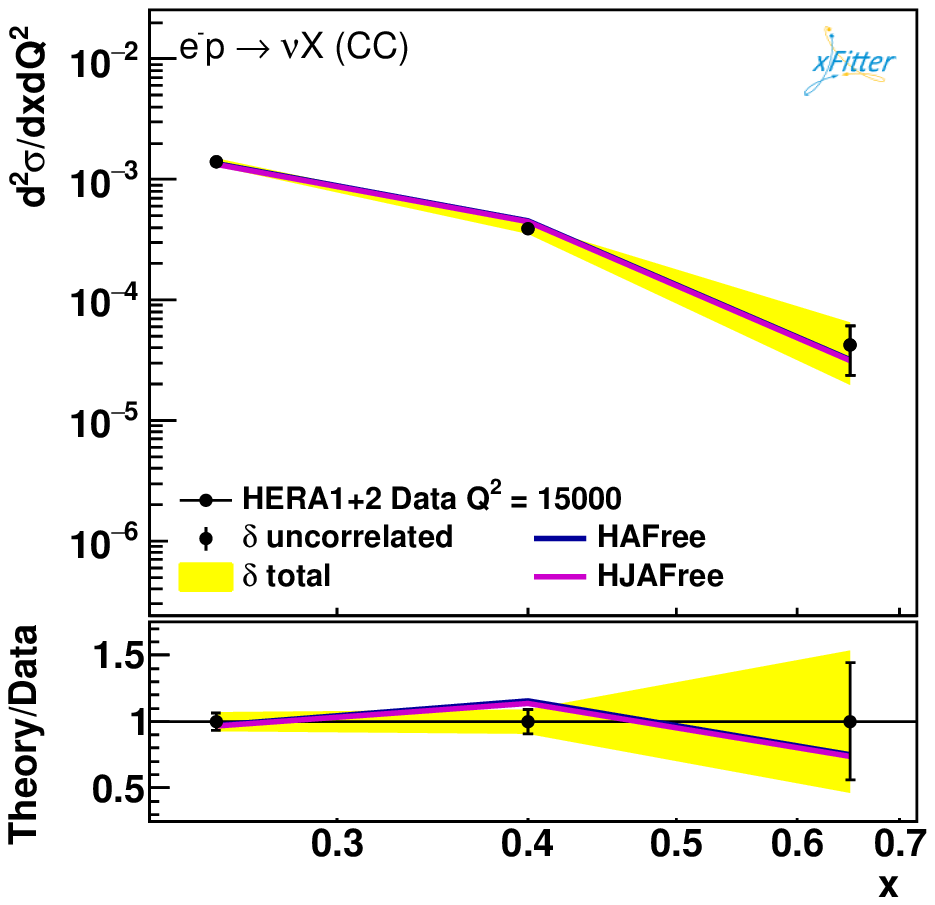}

\caption{The CC deep inelastic ${e^-}p$ scattering double-differential cross sections $\frac{d^2\sigma_{CC}^{e^{\pm} p}}{d{x}dQ^2}$ as a function of $x$ and consistency of data with theory predictions at high $Q^2$.}
\label{fig:2}
\end{figure*}

\begin{figure*}
\includegraphics[width=0.49\textwidth]{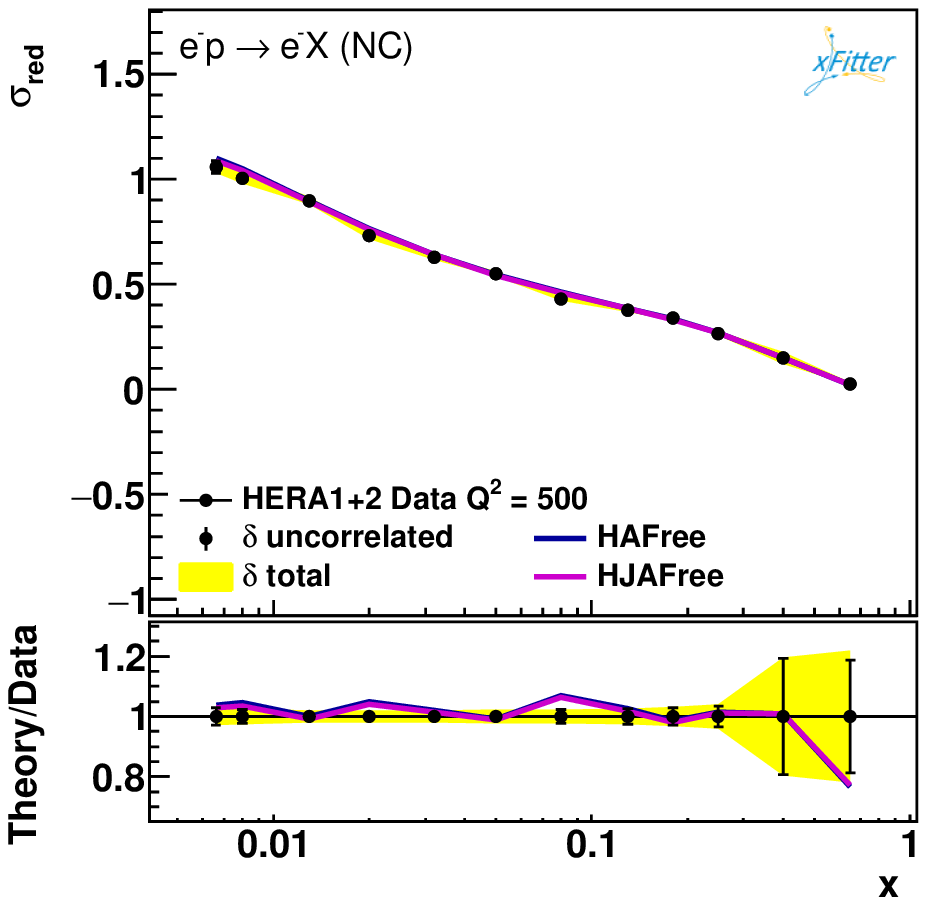}
\includegraphics[width=0.49\textwidth]{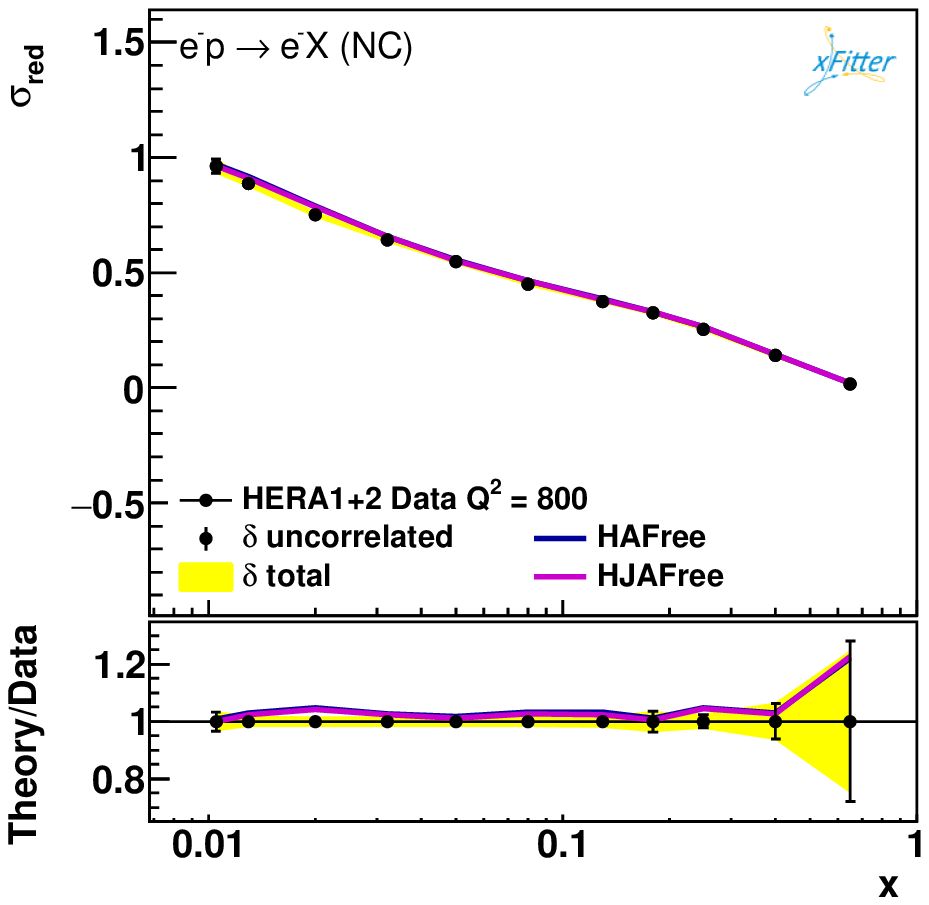}

\caption{The NC deep inelastic ${e^-}p$ scattering cross sections $\sigma_{NC}^{e^{\pm} p}$ as a function of $x$ and consistency of data with theory predictions at low $Q^2$.}
\label{fig:3}
\end{figure*}

\begin{figure*}
\includegraphics[width=0.49\textwidth]{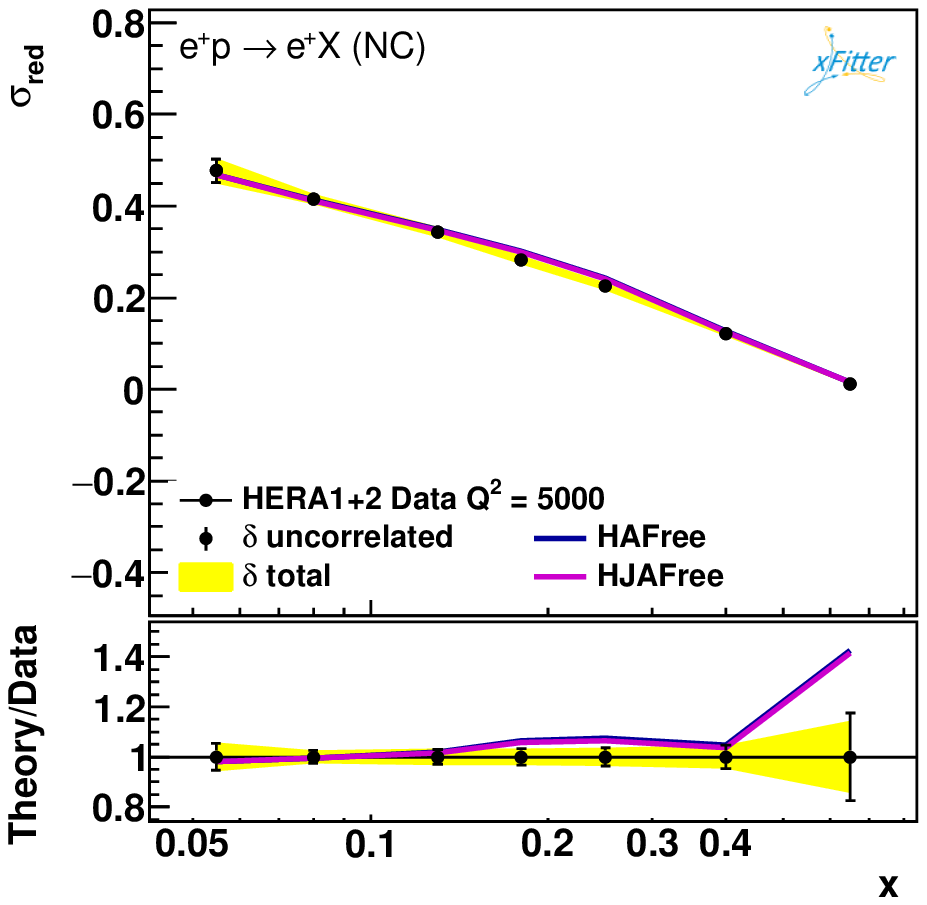}
\includegraphics[width=0.49\textwidth]{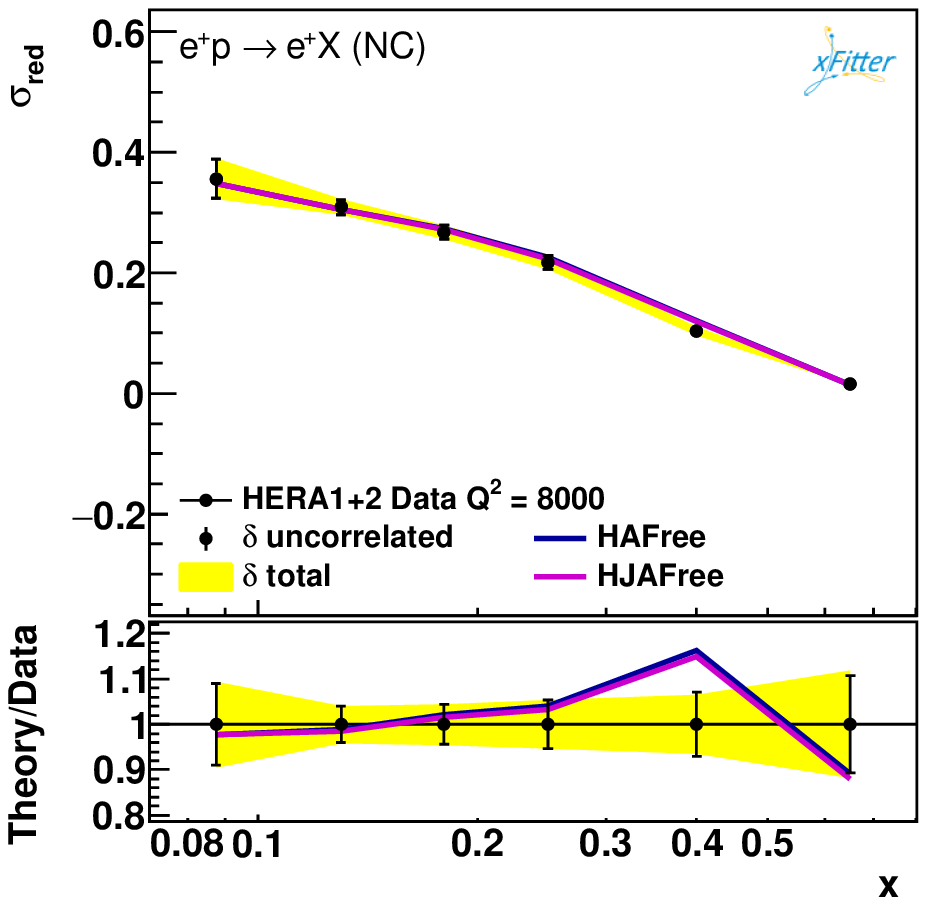}

\caption{The NC deep inelastic ${e^+}p$ scattering cross sections $\sigma_{NC}^{e^{\pm} p}$ as a function of $x$ and consistency of data with theory predictions at high $Q^2$.}
\label{fig:4}
\end{figure*}

This NLO QCD analysis has been perform based on the following theory and data sets:
\begin{itemize}
\item {\bf {HERA I and II combined data}}:
Seven data sets from HERA I and II combined of DIS of $e^{\pm}p$ collision play central role for probing the internal structure of proton and quark-gluon detailed dynamics at centre-of-mass energies of up to $\sqrt{s} \simeq 320\,$GeV~\cite{Abramowicz:2015mha}. 

The reduced neutral current (NC) and inclusive unpolarised charged current (CC) of DIS of $e^{\pm} p$ collision may be expressed in terms of proton structure functions as follows:
\begin{eqnarray}
   \sigma_{r,NC}^{{\pm}}&=&   \frac{d^2\sigma_{NC}^{e^{\pm} p}}{d{x}dQ^2} \frac{Q^4 x}{2\pi \alpha^2 Y_+} = \tilde{F_2} \mp \frac{Y_-}{Y_+} x\tilde{F_3} -\frac{y^2}{Y_+} \tilde{F_{\rm L}}~,                                               
    \label{eq:NC}
\end{eqnarray}
and
\begin{eqnarray}
\sigma_{r,CC}^{\pm} &=&\frac{d^2\sigma_{CC}^{e^{\pm} p}}{d{x}dQ^2} \frac{Q^4 x}{2\pi \alpha^2 Y_+} = \frac{Y_{+}}{2}  W_2^{\pm} \mp \frac{Y_{-}}{2}x  W_3^{\pm} - \frac{y^2}{2} W_L^{\pm}~~,
\label{eq:CC}
\end{eqnarray}
respectively, where $x$ is the Bjorken variable, $y$ is the inelasticity, $Q^2$ is the negative of four-momentum-transfer squared or boson virtuality,  $Y_{\pm} = 1 \pm (1-y)^2$ and $\alpha$ is the fine-structure constant. More details may be found in Ref.~\citep{Vafaee:2017nze}.  

The kinematical range of cross sections for NC interactions of $e^{\pm}p$ collisions are as follows: boson virtuality $Q^2$: $0.045 \leq Q^2 \leq 50000 $\,GeV$^2$ and Bjorken variable x: $6 \cdot 10^{-7} \leq x \leq 0.65$ at the inelasticity of the intraction $ 0.005 \leq y = Q^2/(sx) \leq 0.95$.
The kinematical range of cross sections for CC interactions of $e^{\pm}p$ collisions are as follows: 
$200 \leq Q^2 \leq 50000 $\,GeV$^2$ and
$1.3 \cdot 10^{-2} \leq x \leq 0.40$  
at values of $y$-inelasticity between $0.037$ and $0.76$.

\item {\bf {H1 normalized inclusive jet data}}:
Six data sets from H1 Collaboration are include five normalized inclusive jet data at high $Q^2$ and only one normalized inclusive jet data at low $Q^2$. Five normalized inclusive jet data at high $Q^2$ are include one normalized DiJets with unfolding and one normalized TriJets with unfolding~\cite{Aaron:2009vs}. 

The normalized jet production cross sections are defined as a ratio of differential inclusive $1$-jet, $2$-jet and 3-jet cross sections to
the differential NC deep inelastic $e^{\pm}p$ scattering cross section ($\frac{\sigma_{\rm jet}}{\sigma_{\rm NC}}$, $\frac{\sigma_{\textnormal{2-jet}}}{\sigma_{\rm NC}}$ and $\frac{\sigma_{\textnormal{3-jet}}}{\sigma_{\rm NC}}$ ) in a given $Q^2$ bin , multiplied by the respective bin width $W$ in the case of a double differential measurement as follows:

The normalized inclusive jet production cross section ($\frac{\sigma_{\rm jet}}{\sigma_{\rm NC}}$) is measured as a function of $Q^2$ and double differentially as a function of $Q^2$ and $P_T$ as follows:
\begin{eqnarray}\label{jcs}
\frac{\sigma_{\rm jet}}{\sigma_{\rm NC}}\left(Q^2,\, P_T\right) \hspace{1pc} &=& 
\frac{ \text{d}^2\sigma_{\rm jet} / \text{d}Q^2\,\text{d}P_{T} }
     { \text{d}\sigma_{\rm NC}   /  \text{d}Q^2}
     \cdot W(P_T)~,
\end{eqnarray}     
where $P_T$ is the transverse jet momentum in the Breit frame. 

The normalized $2$ and $3$-jets production cross sections ($\frac{\sigma_{\textnormal{2-jet}}}{\sigma_{\rm NC}}$ and $\frac{\sigma_{\textnormal{3-jet}}}{\sigma_{\rm NC}}$) are presented as a function of $Q^2$ and double differentially as a function of $Q^2$ and $\left\langle P_T\right\rangle$ as follows:
\begin{eqnarray}\label{2jcs}
\frac{\sigma_{\textnormal{2-jet}}}{\sigma_{\rm NC}}\left(Q^2,\, \left\langle P_T\right\rangle\right) &=& 
\frac{ \text{d}^2\sigma_{\textnormal{2-jet}} / \text{d}Q^2\,\text{d}\left\langle P_T\right\rangle }
     { \text{d}\sigma_{\rm NC}   /  \text{d}Q^2}\cdot W(\left\langle P_T\right\rangle)~,
\end{eqnarray}
where $\left\langle P_T\right\rangle$ is the average transverse momentum of the two leading jets, which in turns defined as follows:
\begin{eqnarray}\label{atm}
\left\langle P_T
\right\rangle = \frac{P_{T}^{\rm jet1}+P_{T}^{\rm jet2}}{2}
\end{eqnarray}

In addition, the $2$-jet cross section is
measured double differentially sometimes as a function of $Q^2$ and $\xi$ as follows:
\begin{eqnarray}
\frac{\sigma_{\textnormal{2-jet}}}{\sigma_{\rm NC}}\left(Q^2,\, \xi \right) &=& 
\frac{ \text{d}^2\sigma_{\textnormal{2-jet}} / \text{d}Q^2\,\text{d}\xi }
     { \text{d}\sigma_{\rm NC}   /  \text{d}Q^2}\cdot W(\xi)~,
\end{eqnarray}
where $\xi$ is proton momentum fraction.
It should be noted that the $3$-jet cross section is normalized to the $2$-jet cross section as
function of $Q^2$. 

The kinematical range of cross sections for NC interactions of $e^{\pm}p$ collisions are as follows:
\begin{center}
$150 < Q^2 < 15000$~GeV$^2$~~~{\rm and}~~~$0.2 < y=Q^2/(s\,x) < 0.7$\,,
\end{center}
where as before $Q^2$ is boson virtuality and $y$ is inelasticity of the interaction. The NC deep inelastic of $e^{\pm}p$ collision of event and jet selection are provided for $E_{e^\pm} = 27.6$~GeV with protons of energy $E_p = 920$~GeV, providing a centre-of-mass energy $\sqrt{s}=319$~GeV. 

\item {\bf {ZEUS inclusive jet data}}:
Data sets from ZEUS Collaboration are include three normalized inclusive jet data at high $Q^2$, which one of them is DiJets with unfolding and other two data sets are $1$- jet~\cite{Chekanov:2002be}. These data were taken using ZEUS detector at HERA and inclusive jet differential cross sections have been published in NC of deep inelastic $e^+p$ scattering reaction 
$e^+p\rightarrow e^+\gamma p$ at a centre-of-mass energy of $\sqrt{s}=300$~GeV and correspond to an integrated luminosity of $38.6 \pm 0.6~$pb$^{-1}$.

The kinematical range of cross sections for NC interactions of $e^{+}p$ collisions are as follows:
\begin{center}
$Q^2> 125$~GeV$^2$ and $-0.7\leq\cos \gamma \leq 0.5$~,
\end{center}
where as usual $Q^2$ is known as boson virtuality.  

\end{itemize}
In Figs.~\ref{fig:1}-\ref{fig:4}, we show consistency of theory of NC and CC deep inelastic ${e^+}p$ scattering cross sections and double-differential cross sections with HERA I and II combined experimental data sets based on our NLO DIS analysis.

\begin{figure*}
\includegraphics[width=0.23\textwidth]{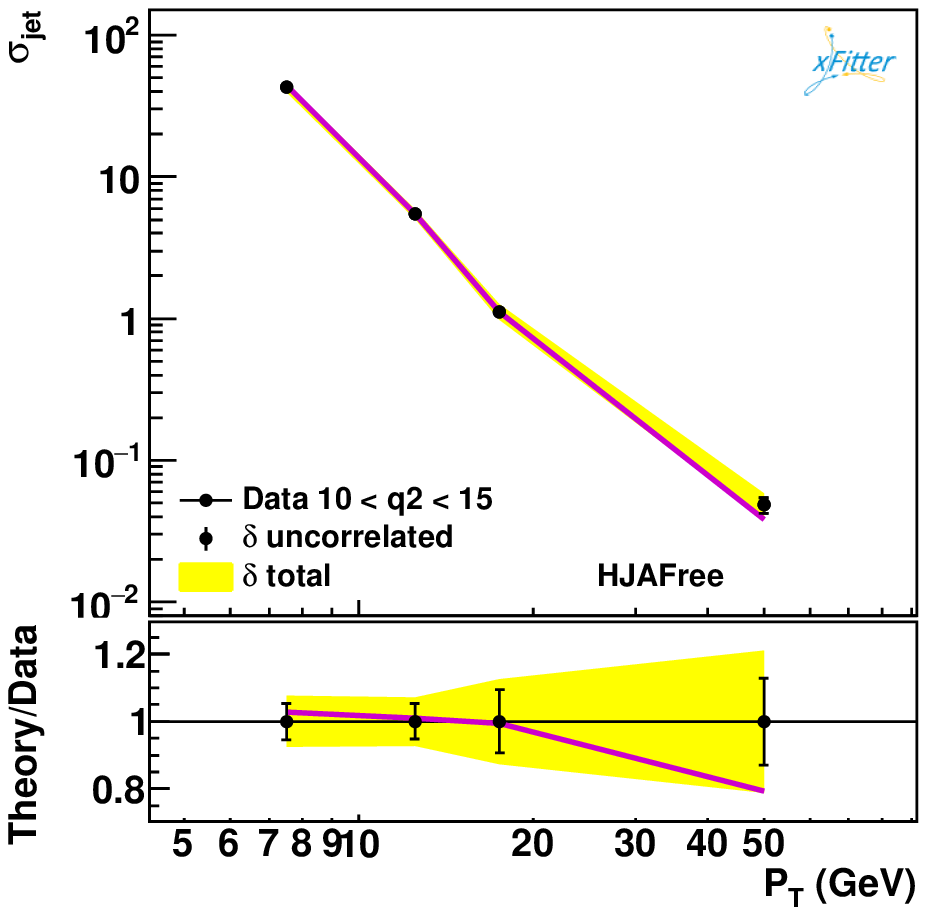}
\includegraphics[width=0.23\textwidth]{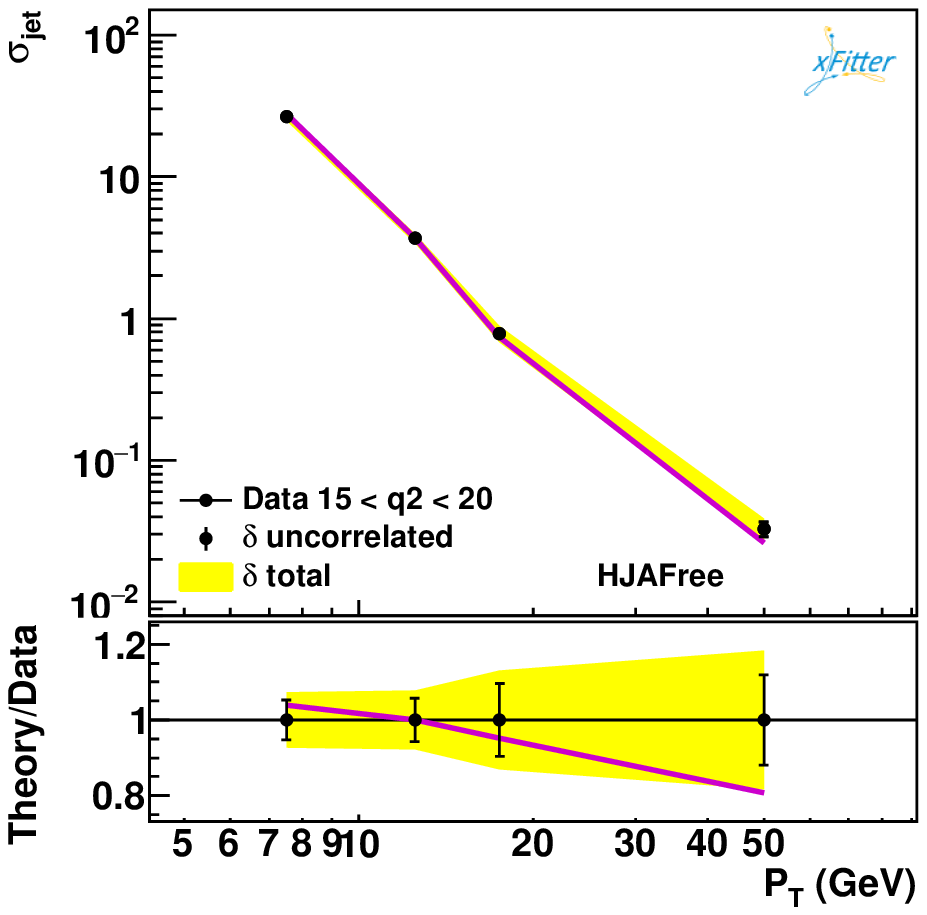}
\includegraphics[width=0.23\textwidth]{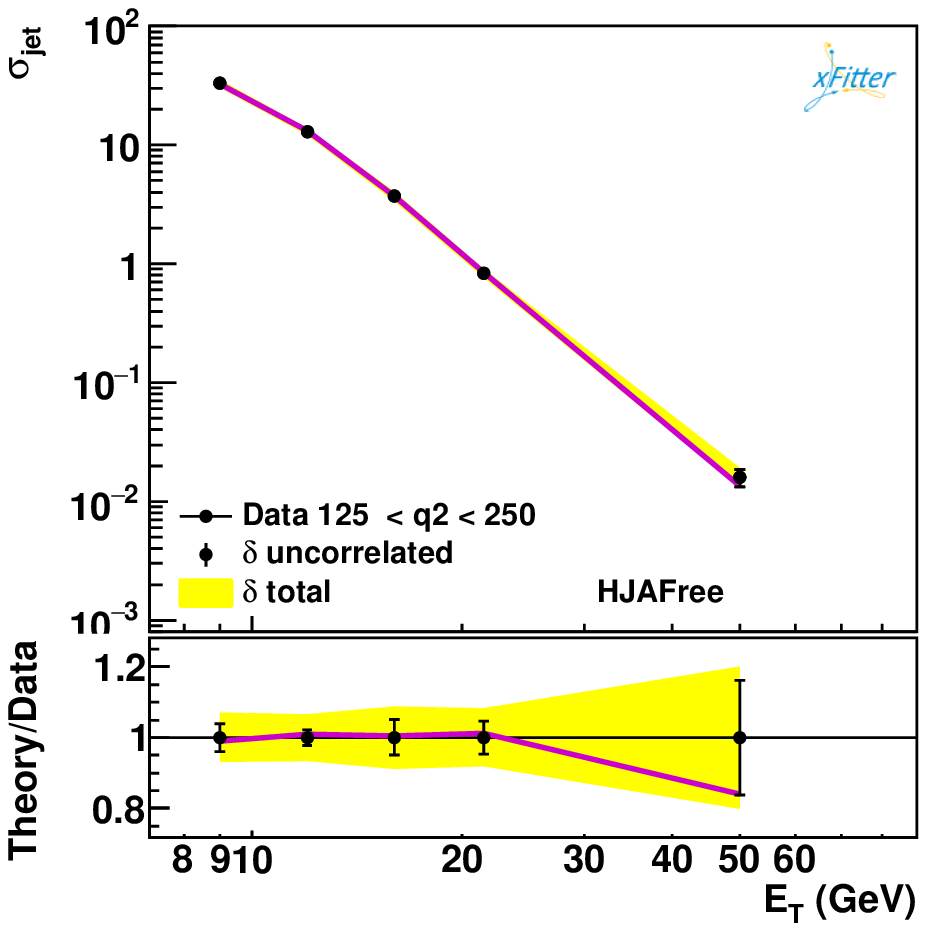}
\includegraphics[width=0.23\textwidth]{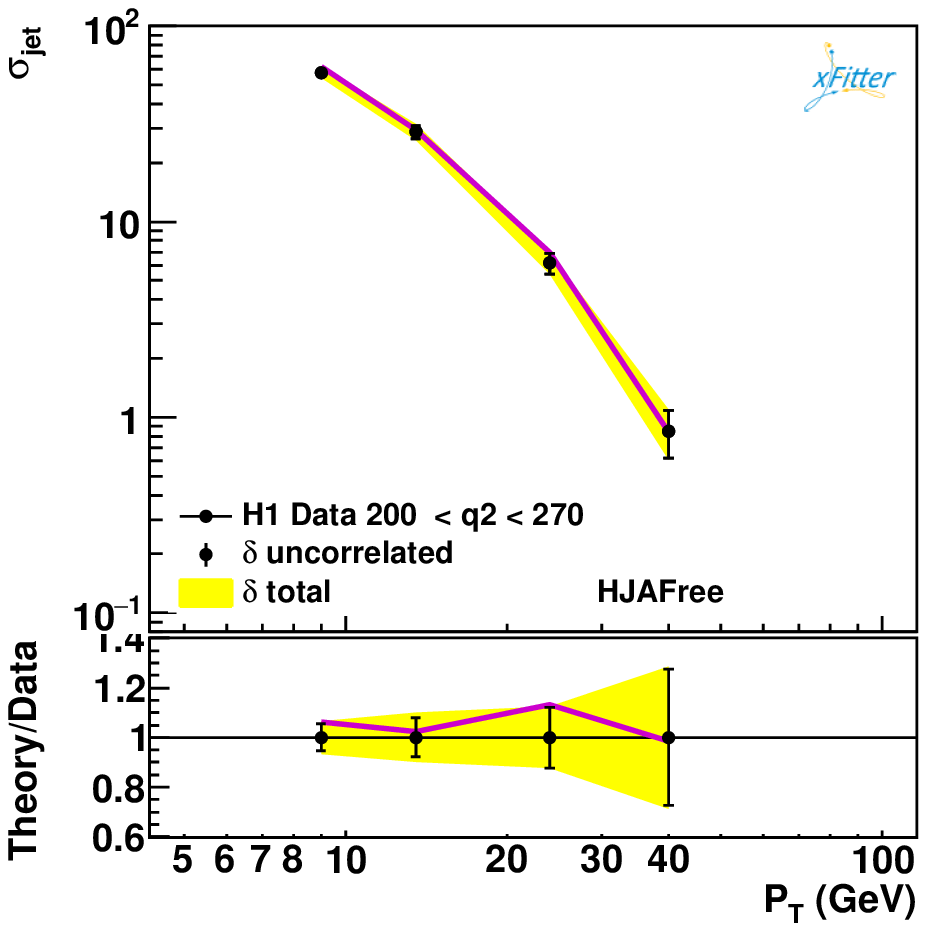}

\includegraphics[width=0.23\textwidth]{./figs/csj1}
\includegraphics[width=0.23\textwidth]{./figs/csj2}
\includegraphics[width=0.23\textwidth]{./figs/csj3}
\includegraphics[width=0.23\textwidth]{./figs/csj4}

\caption{The NC deep inelastic ${e^\pm}p$ scattering jet production cross sections $\sigma_{\rm jet}$ as a function of $x$ and consistency of data with theory predictions at low and high values of $Q^2$.}
\label{fig:5}
\end{figure*}

\begin{figure*}
\includegraphics[width=0.23\textwidth]{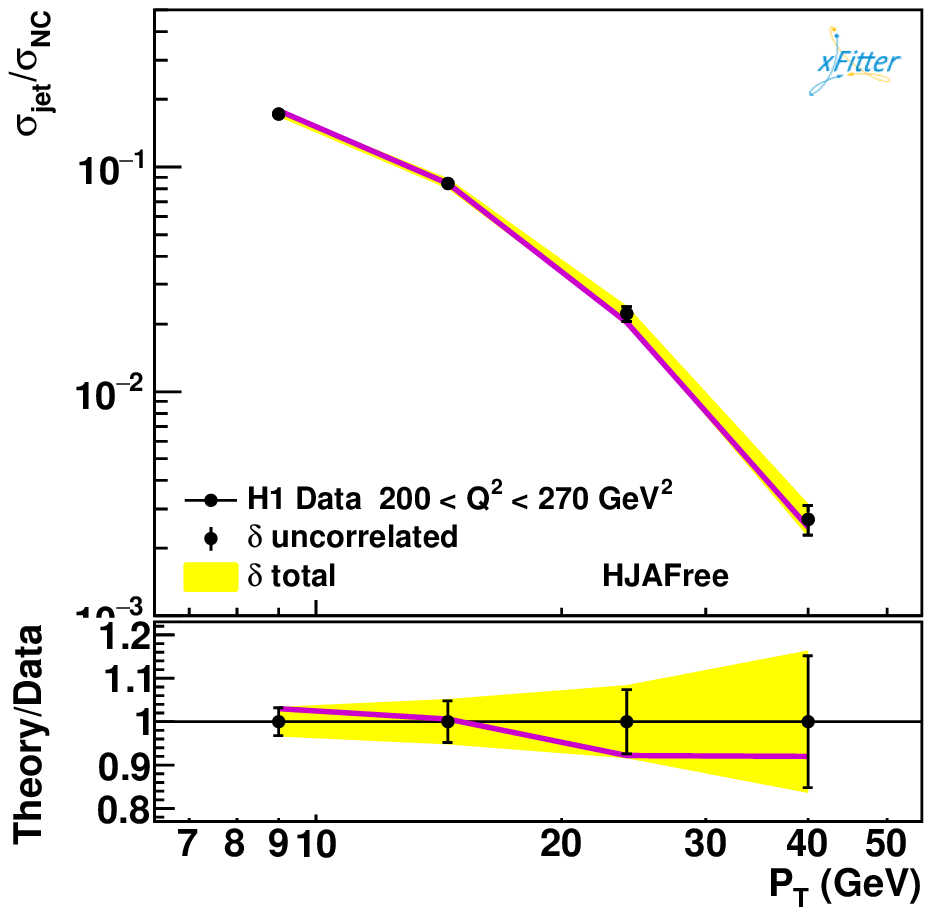}
\includegraphics[width=0.23\textwidth]{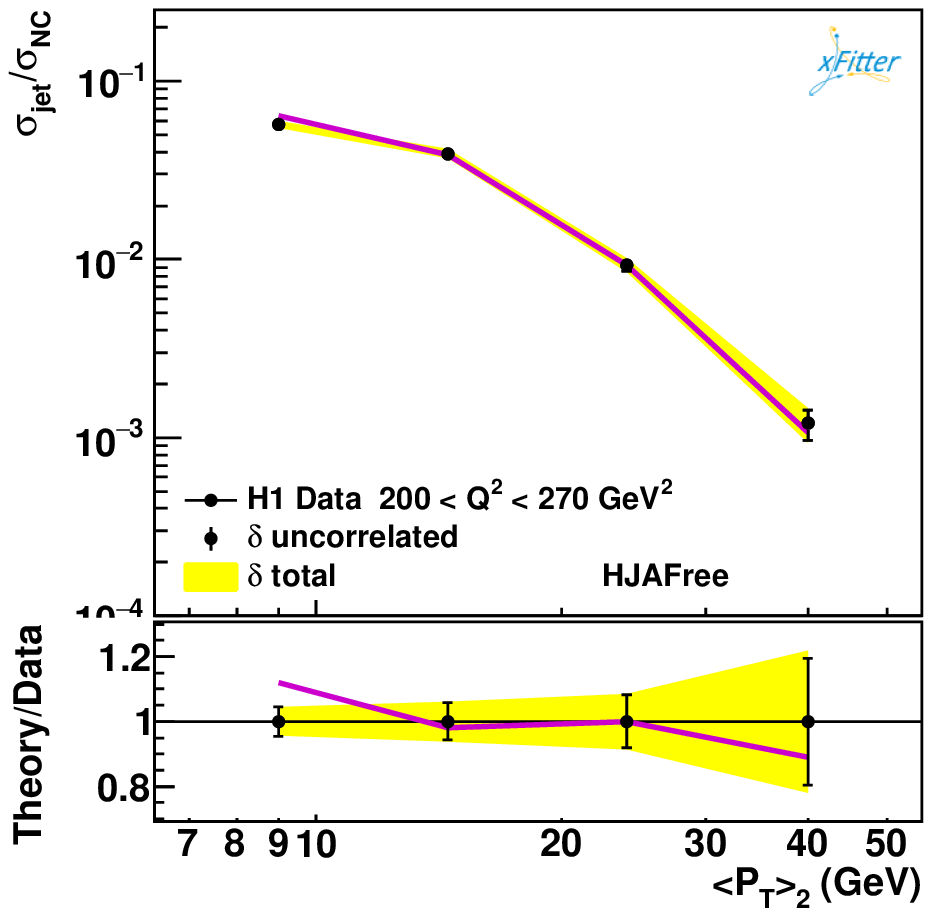}
\includegraphics[width=0.23\textwidth]{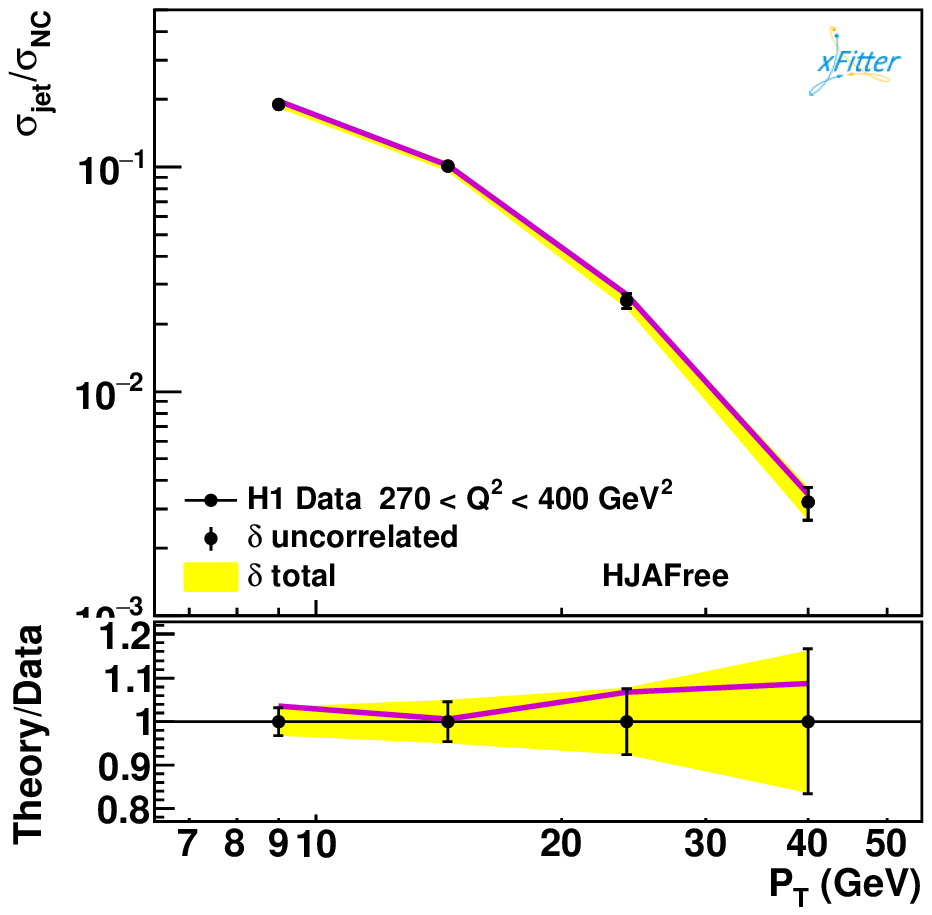}
\includegraphics[width=0.23\textwidth]{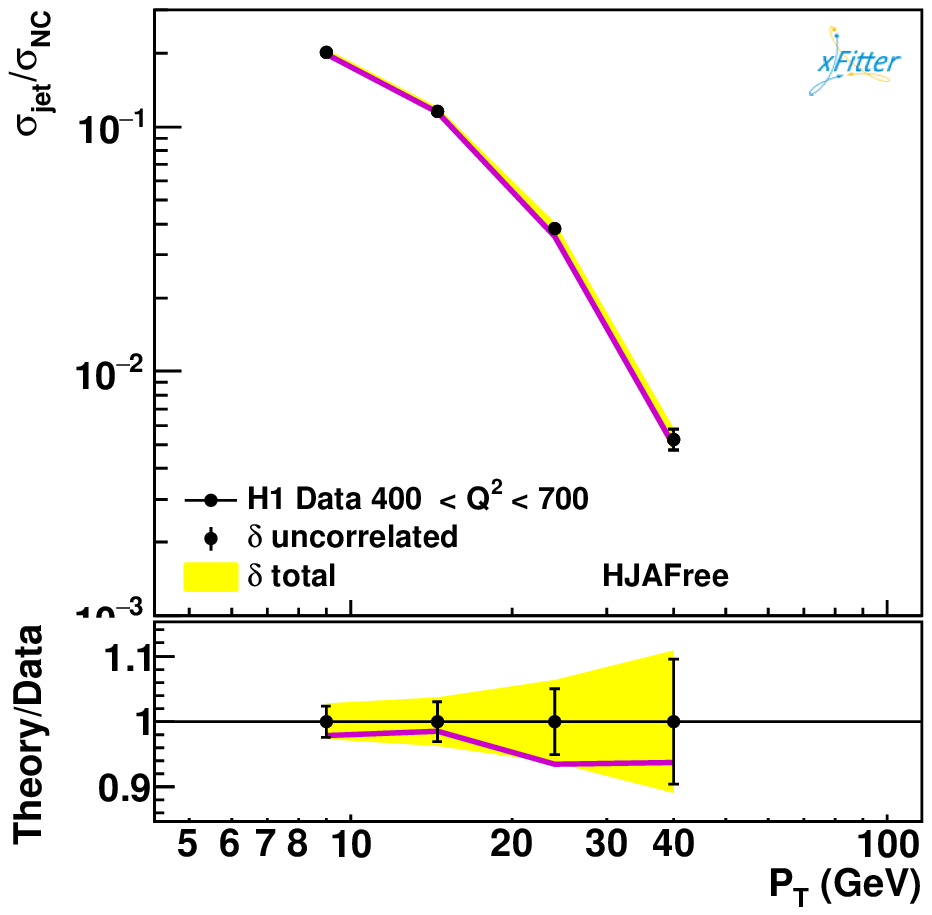}

\includegraphics[width=0.23\textwidth]{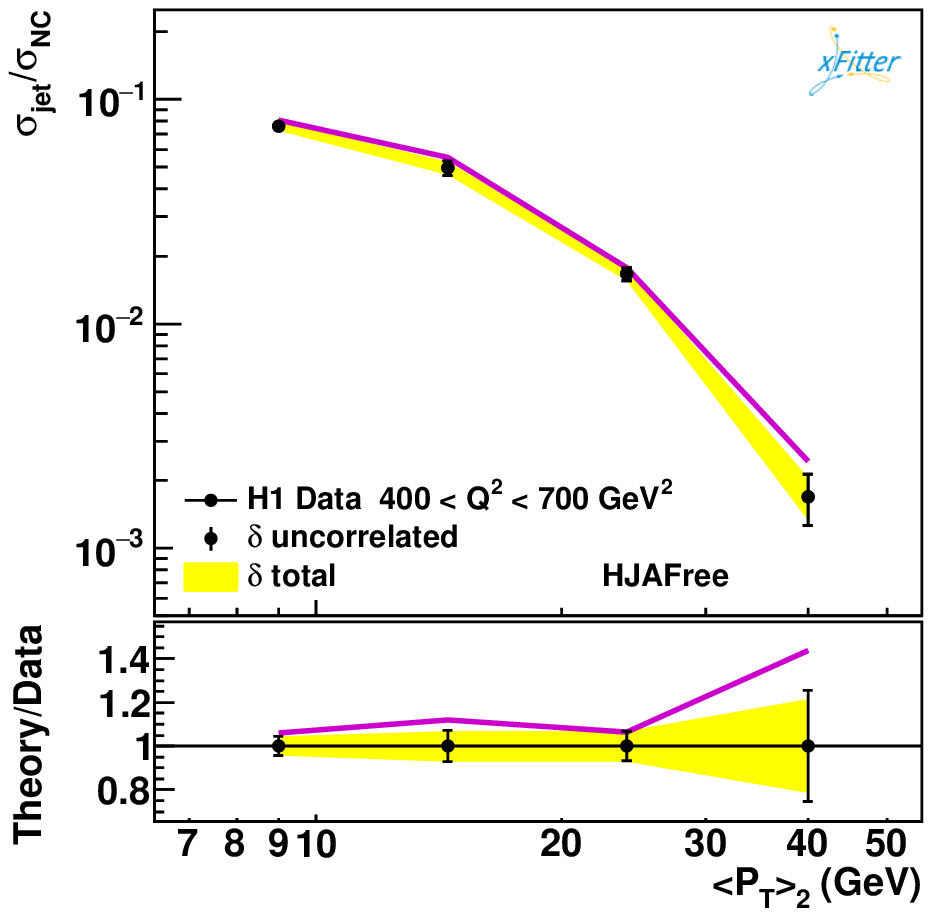}
\includegraphics[width=0.23\textwidth]{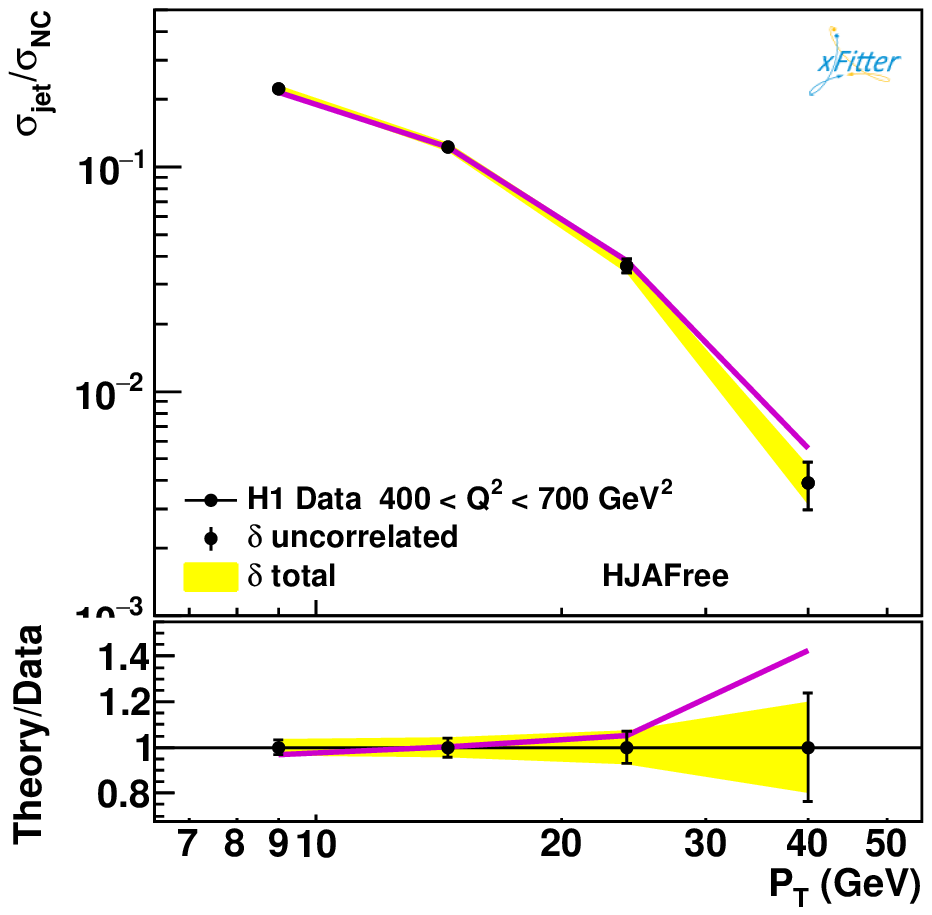}
\includegraphics[width=0.23\textwidth]{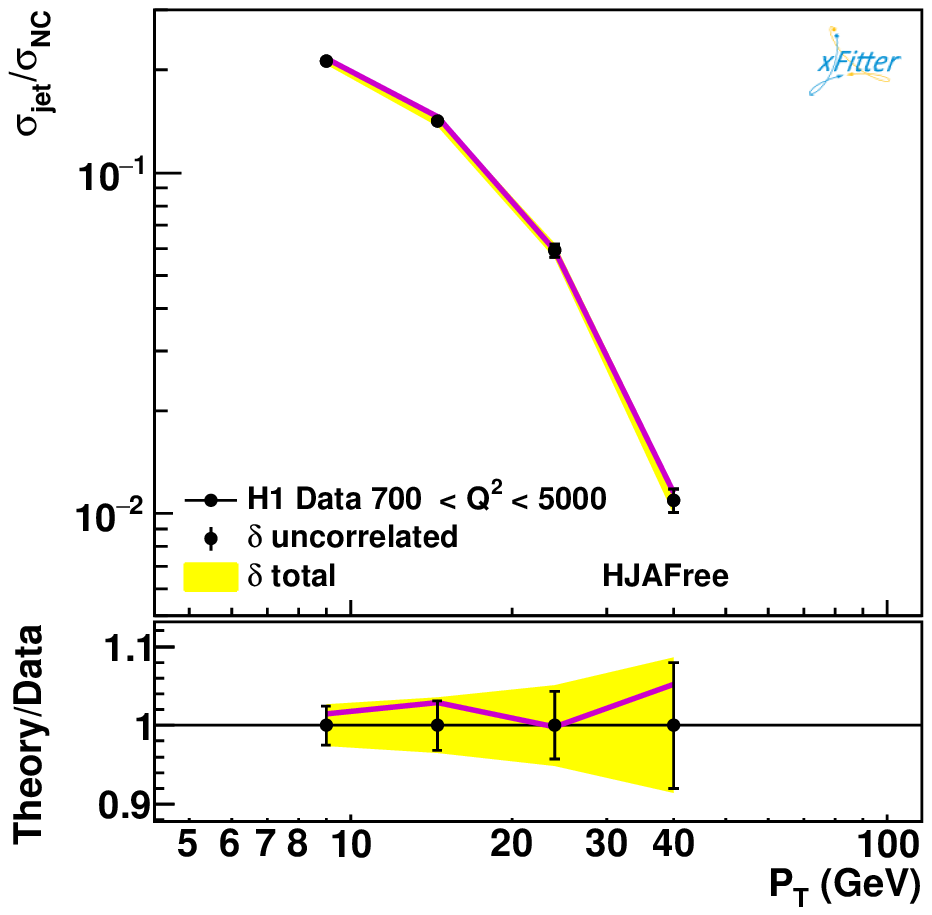}
\includegraphics[width=0.23\textwidth]{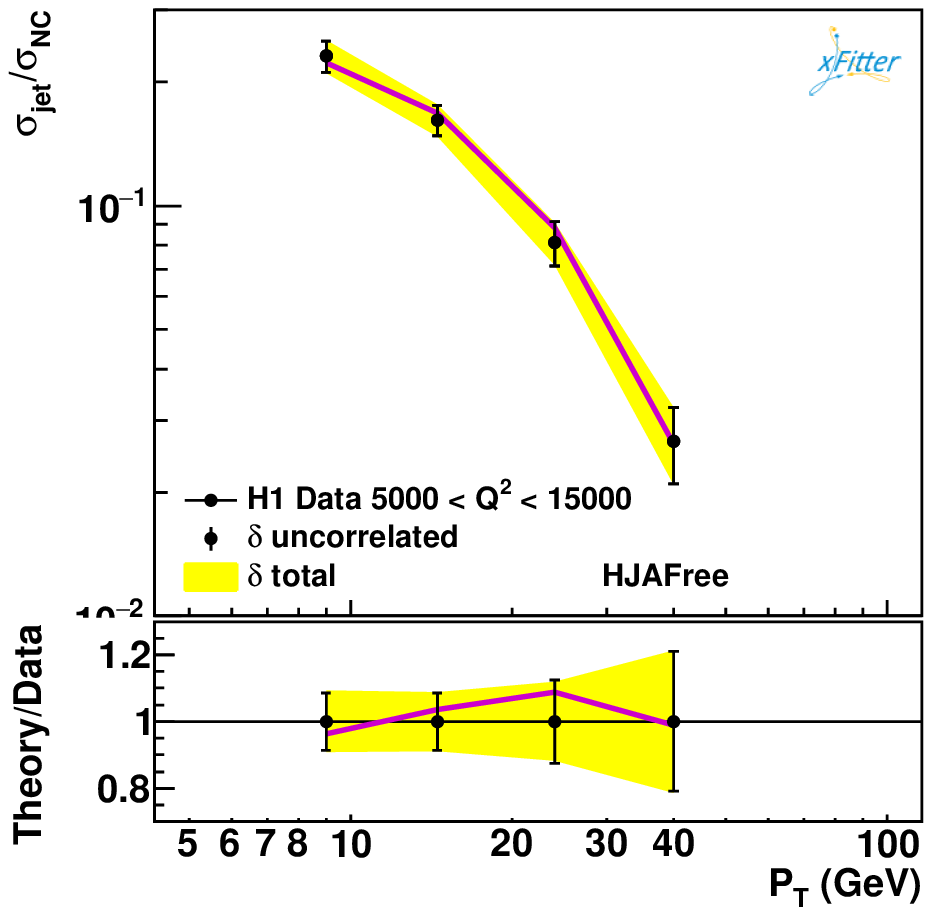}

\caption{The NC deep inelastic ${e^\pm}p$ scattering normalized inclusive jet production cross section $\frac{\sigma_{\rm jet}}{\sigma_{\rm NC}}$ as a function of $x$ and consistency of data with theory predictions at low and high values of $Q^2$.}
\label{fig:6}
\end{figure*}

In Figs.~\ref{fig:5}-\ref{fig:6}, we show consistency of NC deep inelastic ${e^\pm}p$ scattering inclusive cross sections and inclusive normalized cross sections theory with inclusive H1 and ZEUS jet, DiJets and TriJets production cross sections experimental data sets based on our NLO DIS analysis.

It should be noted that in this NLO QCD analysis we perform four different fits titled: HAFixed, HJAFixed, HAFree and HJAFree so that in the throughout of this paper the words HAFixed, HJAFixed, HAFree and HJAFree refer as follows:
\begin{itemize}
\item {\bf HAFixed:} HERA I and II combined data with fixed $\alpha_s(M^2_Z)$.
\item {\bf HJAFixed:} HERA I and II combined data plus H1 plus ZEUS inclusive jet production data sets with fixed $\alpha_s(M^2_Z)$.
\item {\bf HAFree:} HERA I and II combined data with free $\alpha_s(M^2_Z)$.
\item {\bf HJAFree:} HERA I and II combined data plus H1 plus ZEUS inclusive jet production data sets with free $\alpha_s(M^2_Z)$.
\end{itemize}

\section{\label{fit}Fitting and QCD set-up}
\begin{itemize}

\item {\bf PDFs:}  To extract simultaneously PDFs and the strong coupling $\alpha_s^{{\rm NLO}}(M^2_Z)$ of this NLO QCD analysis, we parametrize the proton PDFs based on standard HERAPDF functional form as follows:

\begin{equation}
 xf(x) = A x^{B} (1-x)^{C} (1 + D x + E x^2)~~,
\label{eqn:pdf}
\end{equation}
at the initial scale of the QCD evolution $Q^2_0= 1.9$ GeV$^2$.

 As we mentioned before, in this NLO QCD analysis we perform four different fits with fixed and free $\alpha_s(M^2_Z)$ to investigate the pure impact of inclusion of the inclusive jet production data to the HERA I and II combined NC and CC deep $e^{\pm}p$ scattering cross section data sets and performing an accurate simultaneous determination of $\alpha_s^{{\rm NLO}}(M^2_Z)$ and the gluon distribution. 
 
Based on the standard HERAPDF functional form in Eq.~(\ref{eqn:pdf}) and MINUIT file of this NLO QCD analysis, there are $14$ and $15$ unknown fit parameters corresponding to fits with fixed and free $\alpha_s(M^2_Z)$, respectively which they should be determined by fits to experimental data. 

\begin{table}[h]
\begin{center}
\begin{tabular}{|l|c|c|c|c|}
\hline
\hline
\multicolumn{5}{|c|}{ {Determination of $14$ and $15$ fit parameters corresponding to fits with fixed and free $\alpha_s(M^2_Z)$} }    \\ \hline
 {Parameter} & {$~~~~$HAFixed$~~~~$} & { $~~~$HJAFixed$~~~$} & {HAFree} & {HJAFree} \\ \hline
  ${B_{u_v}}$ & $0.723 \pm 0.046$& $0.696 \pm 0.036$& $0.710 \pm 0.046$& $0.712 \pm 0.039$  \\ 
  ${C_{u_v}}$ & $4.841 \pm 0.087$& $4.818 \pm 0.087$& $4.88 \pm 0.12$& $4.785 \pm 0.088$  \\ 
  $E_{u_v}$ & $13.6 \pm 2.6$& $14.2 \pm 2.3$& $14.0 \pm 2.7$& $14.5 \pm 2.4$  \\ \hline
  ${B_{d_v}}$ & $0.818 \pm 0.095$& $0.852 \pm 0.094$& $0.809 \pm 0.093$& $0.818 \pm 0.097$  \\ 
  $C_{d_v}$ & $4.16 \pm 0.40$& $4.29 \pm 0.41$& $4.18 \pm 0.44$& $3.99 \pm 0.40$ \\ \hline
  $C_{\bar{U}}$ & $8.91 \pm 0.81$& $7.48 \pm 0.76$& $9.1 \pm 1.2$& $7.63 \pm 0.77$  \\ 
  $D_{\bar{U}}$ & $17.7 \pm 3.3$& $10.6 \pm 2.6$& $18.6 \pm 4.5$& $11.1 \pm 2.8$ \\ 
  $A_{\bar{D}}$ & $0.158 \pm 0.011$& $0.170 \pm 0.011$& $0.160 \pm 0.014$& $0.166 \pm 0.011$  \\  
  $B_{\bar{D}}$ & $-0.1682 \pm 0.0082$& $-0.1589 \pm 0.0077$& $-0.166 \pm 0.013$& $-0.1645 \pm 0.0080$  \\ 
  $C_{\bar{D}}$ & $4.2 \pm 1.3$& $7.8 \pm 2.0$& $4.4 \pm 1.3$& $5.4 \pm 1.6$ \\ \hline
  $B_g$ & $-0.11 \pm 0.16$& $-0.07 \pm 0.16$& $-0.13 \pm 0.22$& $-0.03 \pm 0.19$  \\ 
  $C_g$ & $11.2 \pm 1.6$& $7.56 \pm 0.64$& $11.9 \pm 4.1$& $7.44 \pm 0.75$  \\ 
  $A_g'$ & $2.1 \pm 1.4$& $0.61 \pm 0.60$& $2.3 \pm 2.6$& $0.62 \pm 0.72$  \\ 
  ${B_g'}$ & $-0.206 \pm 0.078$& $-0.266 \pm 0.043$& $-0.21 \pm 0.11$& $-0.238 \pm 0.045$  \\ 
  \hline 
  {$\alpha_s^{{\rm NLO}}(M^2_Z)$} & $\textcolor{blue}{ 0.1176 }$& $\textcolor{blue}{ 0.1176 }$& $0.1160 \pm 0.0049$& $0.12041 \pm 0.00086$ \\ \hline
\hline
    \end{tabular}
\vspace{-0.0cm}
\caption{\label{tab:par}{ {Numerical values of $14$ and $15$ free central parameters and their uncertainties corresponding to four different HAFixed, HJAFixed, HAFree and HJAFree fits, respectively.}}}
\vspace{-0.4cm}
\end{center}
\end{table}  

 In Table~\ref{tab:par}, we present numerical values of $14$ and $15$ free central parameters and their uncertainties corresponding to fits with fixed and free strong coupling $\alpha_s(M^2_Z)$, respectively. 

\item {\bf Fitting:} As we mentioned, in this NLO QCD analysis we use three different data sets as follos:
\begin{enumerate}
\item Seven sets of HERA I and II combined data, as the central data sets for probing the internal structure of proton as a whole~\cite{Abramowicz:2015mha}.

\item Six data sets of H1 normalized inclusive jet production data~\cite{Aaron:2009vs}.

\item Three data sets of ZEUS inclusive jet production data~\cite{Chekanov:2002be}.
\end{enumerate}

\begin{table}[h]
\begin{center}
\begin{tabular}{|l|c|c|c|c|}
\hline
\hline
\multicolumn{5}{|c|}{ {Three different data sets of this NLO QCD analysis} }    \\ \hline
 {Experiment} & {$~~~~$HAFixed$~~~~$} & { $~~~$HJAFixed$~~~$} & {$~~~$HAFree$~~~$} & {$~~~$HJAFree$~~~$} \\ \hline
  HERA I+II CC $e^{+}p$~\cite{Abramowicz:2015mha} & 44 / 39& 49 / 39& 45 / 39& 45 / 39 \\ 
  HERA I+II CC $e^{-}p$~\cite{Abramowicz:2015mha} & 49 / 42& 49 / 42& 49 / 42& 50 / 42 \\ 
  HERA I+II NC $e^{-}p$~\cite{Abramowicz:2015mha} & 221 / 159& 221 / 159& 222 / 159& 221 / 159 \\ 
  HERA I+II NC $e^{+}p$ 460~\cite{Abramowicz:2015mha} & 208 / 204& 212 / 204& 209 / 204& 210 / 204 \\ 
  HERA I+II NC $e^{+}p$ 575~\cite{Abramowicz:2015mha} & 213 / 254& 218 / 254& 213 / 254& 216 / 254 \\
  HERA I+II NC $e^{+}p$ 820~\cite{Abramowicz:2015mha} &  66 / 70& 68 / 70& 66 / 70& 67 / 70 \\ 
  HERA I+II NC $e^{+}p$ 920~\cite{Abramowicz:2015mha} & 422 / 377& 440 / 377& 422 / 377& 435 / 377 \\ \hline
  H1 Low $Q^2$ Inclusive Jet Data~\cite{Aaron:2009vs} & - & 24 / 28& - & 24 / 28 \\ 
  H1 Inclusive Jet Data~\cite{Aaron:2009vs} & - & 12 / 24& - & 12 / 24 \\
  H1 Normalized Inclusive Jet Data~\cite{Aaron:2009vs} & - & 12 / 24& - & 12 / 24  \\
  H1 Normalized Inclusive Jets with Unfolding~\cite{Aaron:2009vs} & - & 23 / 24& - & 21 / 24 \\
  H1 Normalized DiJets with Unfolding~\cite{Aaron:2009vs} & - & 37 / 24& - & 41 / 24 \\
  H1 Normalized TriJets with Unfolding~\cite{Aaron:2009vs} & - & 13 / 16& - & 9.5 / 16 \\  \hline
  ZEUS Inclusive Jet Data~\cite{Chekanov:2002be} & - & 27 / 30& - & 25 / 30  \\ 
  ZEUS Inclusive Jet Data~\cite{Chekanov:2002be} & - & 26 / 30& - & 25 / 30  \\ 
  ZEUS Inclusive DiJet Data~\cite{Chekanov:2002be} & - & 18 / 22& - & 17 / 22  \\ \hline   
 { Correlated ${\chi^2}$} & 111& 103& 109& 111  \\ \hline
{${\frac{{\chi^2}_{Total}}{dof}}$} & ${\frac{1335}{1131}}$  & ${\frac{1552}{1353}}$ &  ${\frac{1335}{1130}}$ &  ${\frac{1541}{1352}}$ \\ \hline
\hline
    \end{tabular}
\vspace{-0.0cm}
\caption{\label{tab:data}{Experiments, correlated $\chi^2$ and {$\frac{{\chi^2}_{Total}}{dof}$} corresponding to four different HAFixed, HJAFixed, HAFree and HJAFree fits, respectively.}}
\vspace{-0.4cm}
\end{center}
\end{table}

Table~\ref{tab:data} shows three different data sets of this NLO QCD analysis, correlated $\chi^2$ and {$\frac{{\chi^2}_{Total}}{dof}$} corresponding to four different HAFixed, HJAFixed, HAFree and HJAFree fits.

\item {\bf Scaling:} In this NLO QCD analysis, we set the factorisation and renormalisation scales as follows:
\begin{center}
$\mu_{\rm f}^2 = Q^2$~~~ and~~~ $\mu_{\rm r}^2 = {\frac{1}{2}}{(Q^2 + p_{T}^2)}$~,
\end{center}
respectively, where $p_{T}$ is the transverse
momenta. 

\item {\bf Extra Minimisation Parameters:} In fitting with fixed strong coupling, we set the strong coupling to $\alpha_s(M^2_Z) = 0.1176$ and in fitting with free strong coupling, we varied $\alpha_s(M^2_Z)$ in steps of $0.001$. In addition, we fixed the strangeness suppression factor to $f_{s}=0.31$, which has the best consistency with other physical parameters of this NLO QCD analysis~\cite{James:1975dr,Vafaee:2018abd,Shokouhi:2018gie,Vafaee:2018ehy}.

\item {\bf Evolution of PDFs:} To evolve the parametrized PDFs, we use QCDNUM version $17$-$01/15$ as a very fast QCD evolution program using xFitter QCD framework version $2.0.0$ FrozenFrog and start this NLO QCD analysis evolution at starting scale of $Q^{2}_{0}=1.9$~GeV$^{2}$~\cite{Botje:2010ay,xFitter,Sapronov,Vafaee:2017jnt,Vafaee:2016jxl}.
\end{itemize} 

\section{\label{results}Results}

\begin{itemize}

\item {\bf Fit-quality:}

As we know, in a QCD analysis, proton PDFs are extracted via fitting to experimental data sets based on minimization of $\chi^2$-function and accordingly $\frac{{\chi^2}_{Total}}{dof}$, where dof in the denominator refers to degrees of freedom, is a measure of consistency between theory and experiment. Now if we summarize the numerical values of $\frac{{\chi^2}_{Total}}{dof}$ and {$\alpha_s^{{\rm NLO}}(M^2_Z)$} from Tables~\ref{tab:par} and \ref{tab:data} into the Table~\ref{tab:fq} we may conclude the following results:
\begin{enumerate}

\item According to the relative change of $\chi^2$-function which is defined by $\frac{\chi^2_{\rm final}-\chi^2_{\rm initial}}{\chi^2_{\rm initial}}$~, we obtain up to ${\frac{1.180-1.147}{1.180}}\sim 2.8$~\% relative improvement in the quality of the fit for fit with fixed $\alpha_s(M^2_Z)$. In addition for fit with fixed $\alpha_s(M^2_Z)$, this improvement in the quality of the fit is due to pure impact of inclusion of the inclusive jet production data to the HERA I and II combined NC and CC deep $e^{\pm}p$ scattering cross section data sets.   

\item Improvement the quality of the fit for fit with free $\alpha_s(M^2_Z)$ is: $\frac{\chi^2_{\rm final}-\chi^2_{\rm initial}}{\chi^2_{\rm initial}} = {\frac{1.181-1.139}{1.181}}$, which shows up to $\sim 3.6$~\% relative improvement in the quality of the fit for fit with free $\alpha_s(M^2_Z)$.

\item Now if we compare the results for the quality of the fit for fits with fixed and free $\alpha_s(M^2_Z)$, we conclude that the best fit-quality is related to HJAFree fit in our QCD analysis for inclusion of the inclusive jet production data and a simultaneous determination of PDFs and the strong coupling $\alpha_s(M^2_Z)$. In addition, this comparison clearly shows up to $\mid3.6 - 2.8\mid = 0.8$~\% improvement in the quality of the fit is due to pure contribution of the strong coupling $\alpha_s(M^2_Z)$, when it is considered as an extra fit parameter. This results not only show the strong correlation between $\alpha_s(M^2_Z)$ and proton PDFs but also emphasize the central role of the strong coupling $\alpha_s(M^2_Z)$ at pQCD analysis level.   
\end{enumerate}
 
\begin{table}[h]
\begin{center}
\begin{tabular}{|l|c|c|c|c|}
\hline
\hline
 {QCD Analysis} & {{$\frac{{\chi^2}_{Total}}{dof}$}} & {{$\alpha_s^{{\rm NLO}}(M^2_Z)$}}  \\ \hline 
HAFixed & ${\frac{1335}{1131} = 1.180}$ & $\textcolor{blue}{ 0.1176 }$ \\ \hline
HJAFixed & ${\frac{1552}{1353} = 1.147}$ & $\textcolor{blue}{ 0.1176 }$ \\ \hline
HAFree & ${\frac{1335}{1130} = 1.181}$ & $0.1160 \pm 0.0049$ \\ \hline
HJAFree & ${\frac{1541}{1352} = 1.139}$ & $0.12041 \pm 0.00086$ \\ \hline \hline
  \end{tabular}
\vspace{-0.0cm}
\caption{\label{tab:fq}{Comparison {$\frac{{\chi^2}_{Total}}{dof}$} and the QCD fit-quality corresponding to four different HAFixed, HJAFixed, HAFree and HJAFree analysis.}}
\vspace{-0.4cm}
\end{center}
\end{table}

\item {\bf Determination of $\bf {\alpha_s^{{\rm NLO}}(M^2_Z)}$:}
The methodology of this NLO QCD analysis leads to the values of $\alpha_s^{{\rm NLO}}(M^2_Z) = 0.1160 \pm 0.0049$ and $\alpha_s^{{\rm NLO}}(M^2_Z) = 0.12041 \pm 0.00086 $ corresponding to HAFree and HJAFree analysis. These values of $\alpha_s(M^2_Z)$    are in good agreement with the world average and other individual measurements~\cite{Agashe:2014kda}.

\item {\bf Gluon distribution:}

Fig.~\ref{fig:7} shows comparison of the pure impact of the strong coupling $\alpha_s(M^2_Z)$ on the gluon distribution for fits with fixed (purple) and free (blue) $\alpha_s(M^2_Z)$, without and with inclusion of inclusive jet production data corresponding to upper three and lower three diagrams, respectively.

Comparison of the pure impact of the strong coupling $\alpha_s(M^2_Z)$ on the gluon-ratio distribution for fits with fixed (purple) and free (blue) $\alpha_s(M^2_Z)$, without and with inclusion of inclusive jet production data corresponding to upper three and lower three diagrams, respectively is shown in Fig.~\ref{fig:8}~.

In Fig.~\ref{fig:9} we compare the pure impact of inclusion jet production data on the shape of gluon distribution for fits with fixed (upper three diagrams) and free (lower three diagrams) strong coupling $\alpha_s(M^2_Z)$.

Fig.~\ref{fig:10} shows comparison of the pure impact of inclusion jet production data on the shape of gluon-ratio distribution for fits with fixed (upper three diagrams) and free (lower three diagrams) strong coupling $\alpha_s(M^2_Z)$.

\begin{figure*}
\includegraphics[width=0.28\textwidth]{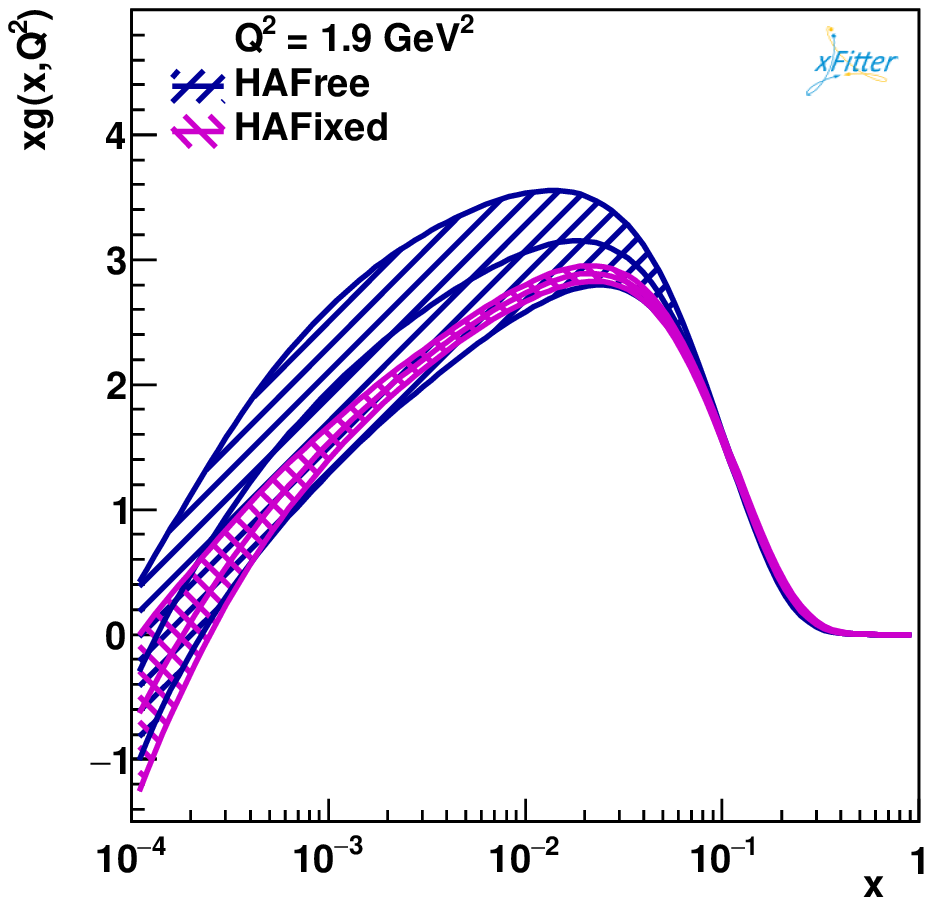}
\includegraphics[width=0.28\textwidth]{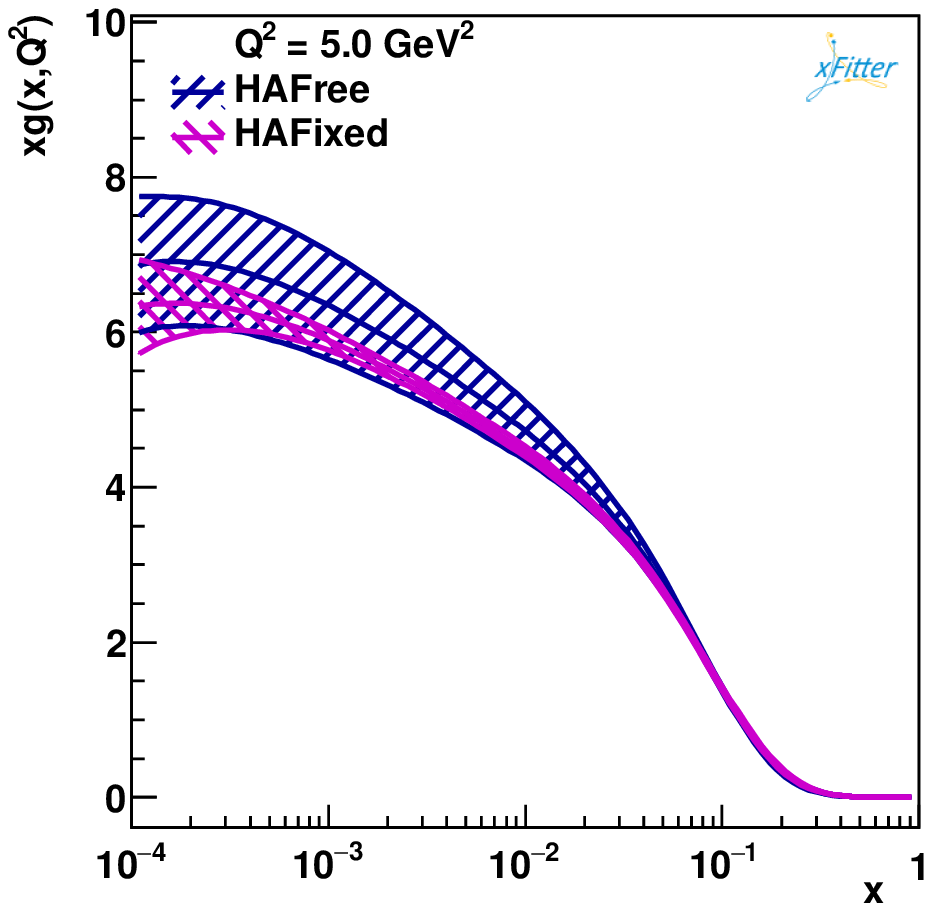}
\includegraphics[width=0.28\textwidth]{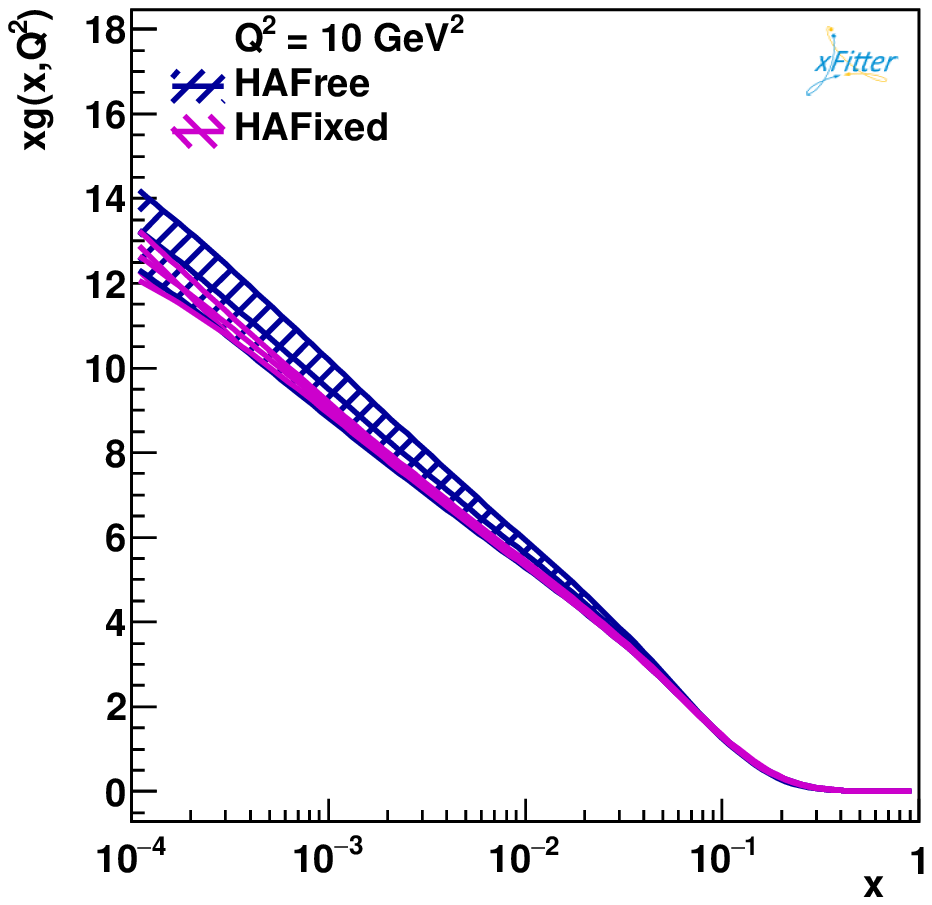}

\includegraphics[width=0.28\textwidth]{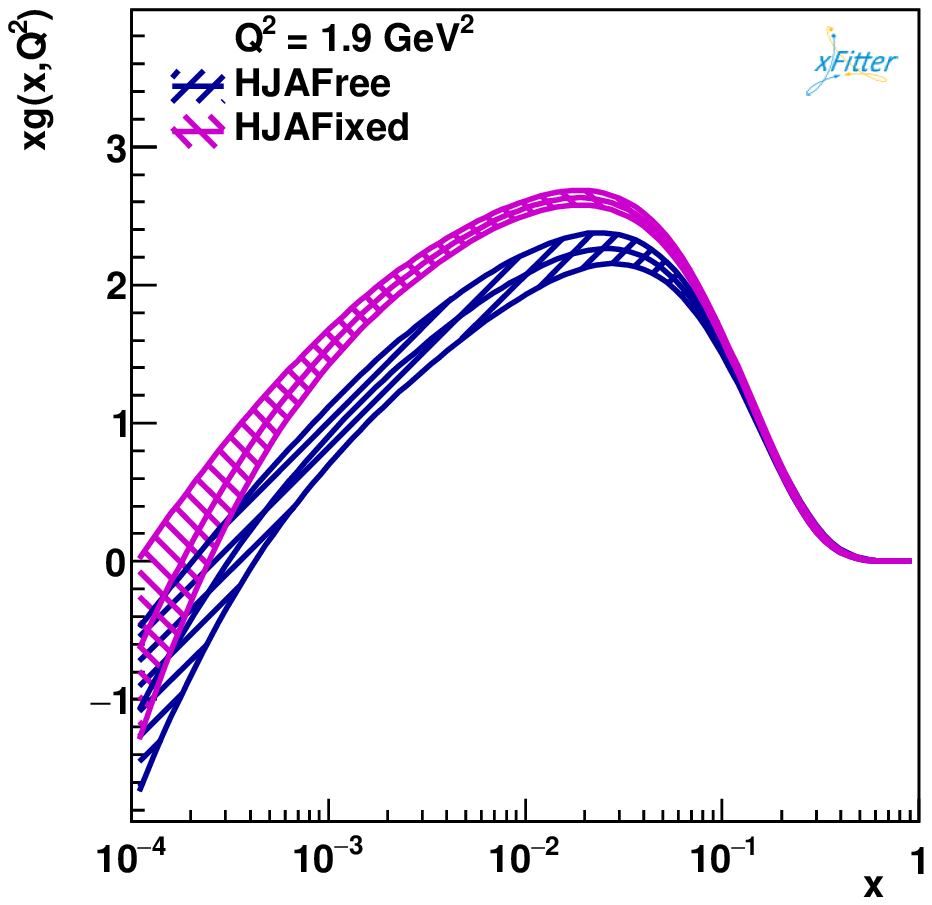}
\includegraphics[width=0.28\textwidth]{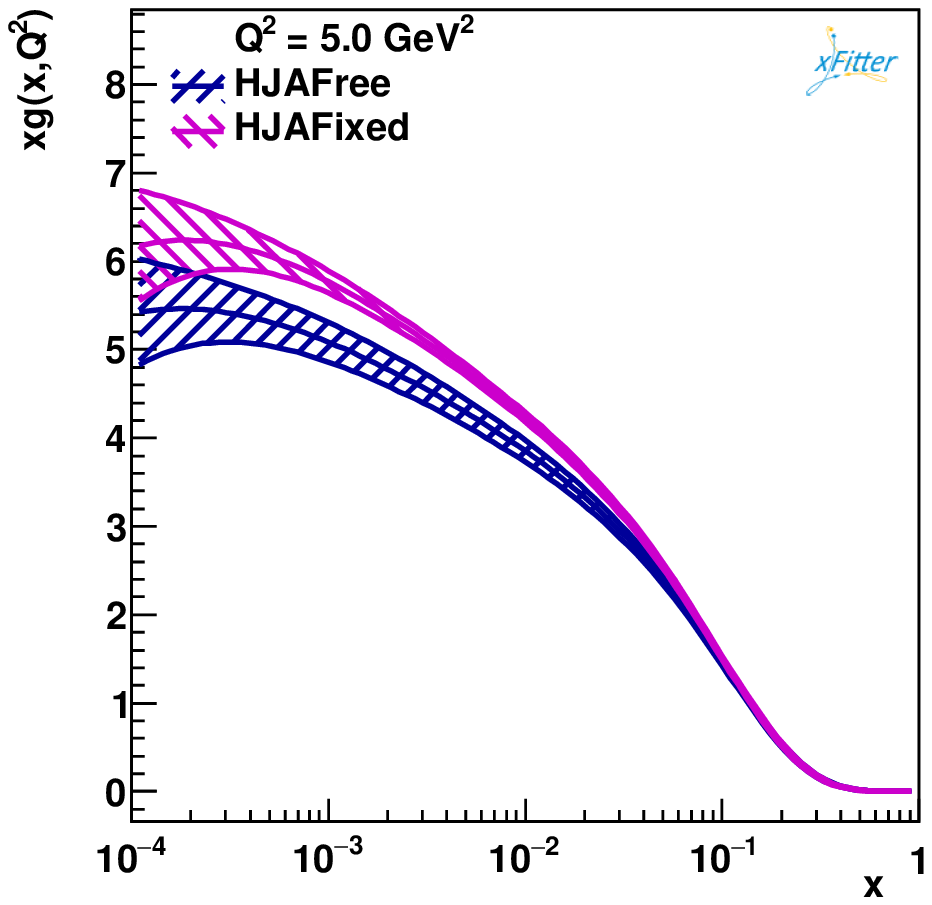}
\includegraphics[width=0.28\textwidth]{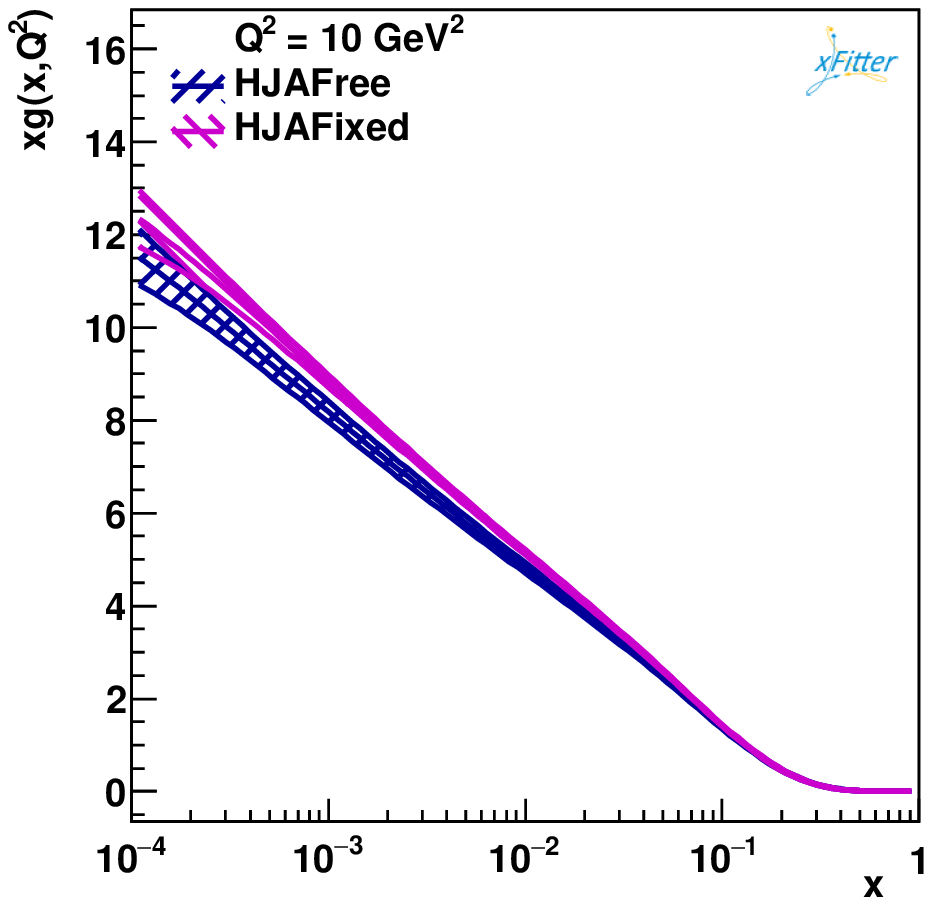}
\caption{Comparison of the pure impact of the strong coupling $\alpha_s(M^2_Z)$ on the gluon distribution for fits with fixed (purple) and free (blue) $\alpha_s(M^2_Z)$, without and with inclusion of inclusive jet production data corresponding to upper three and lower three diagrams, respectively.}
\label{fig:7}
\end{figure*}

\begin{figure*}
\includegraphics[width=0.28\textwidth]{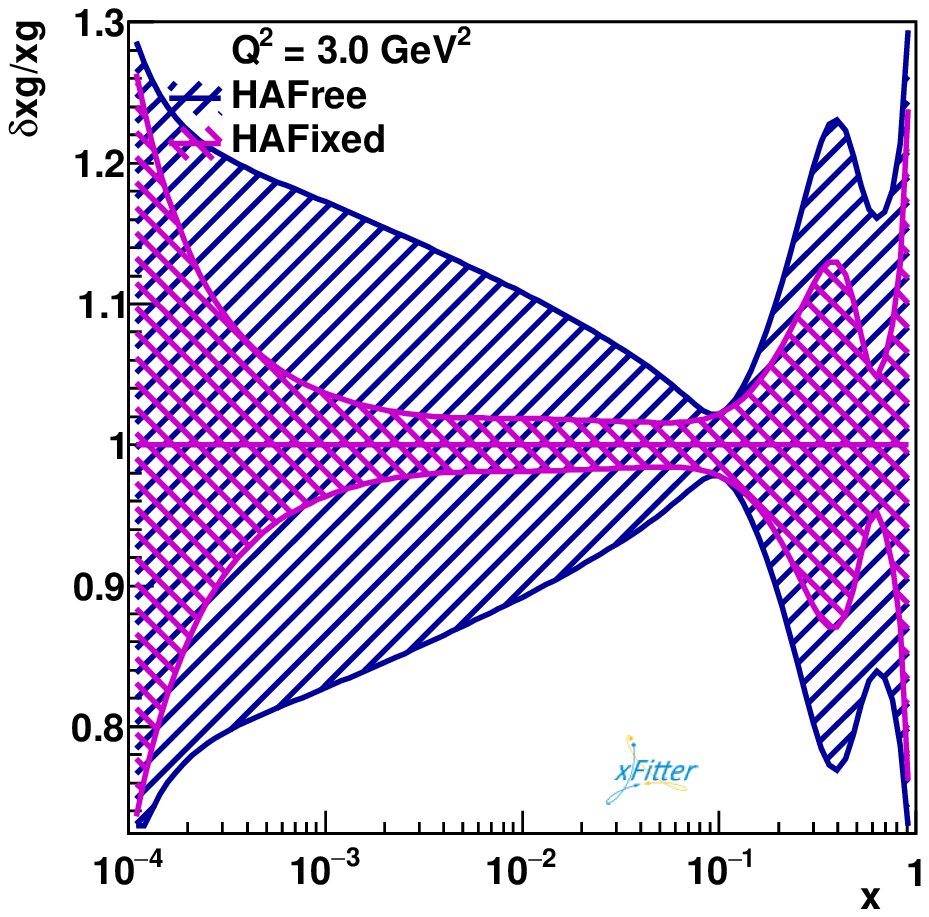}
\includegraphics[width=0.28\textwidth]{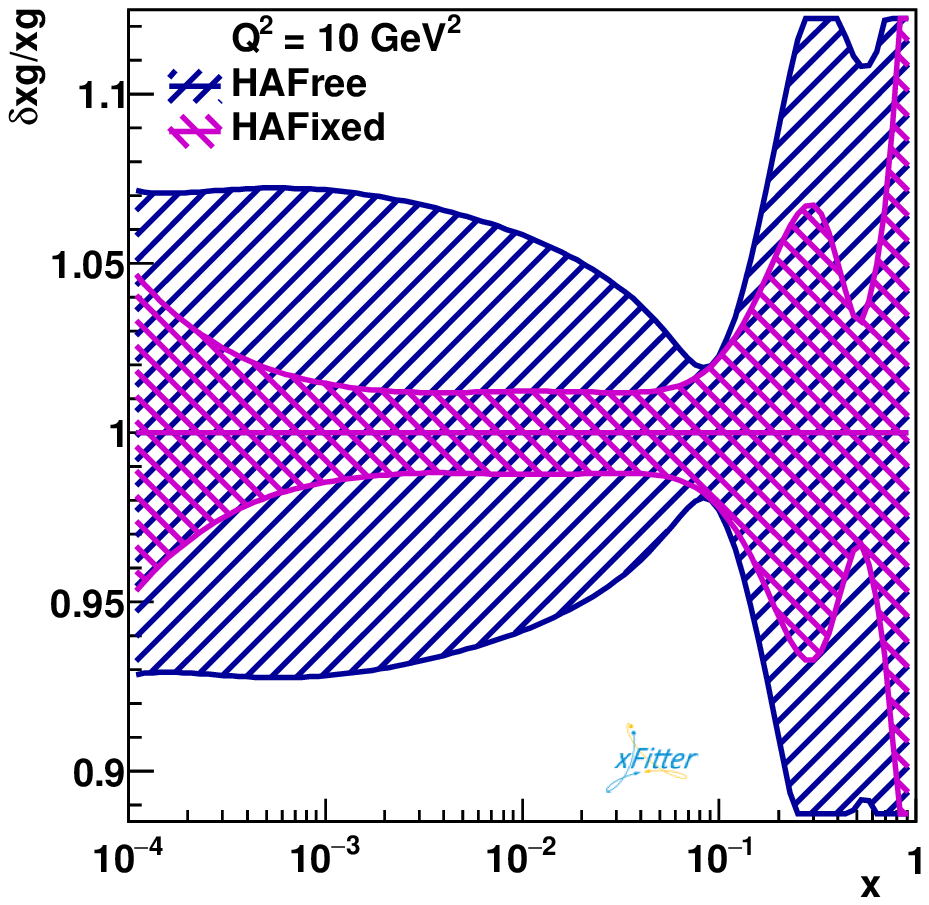}
\includegraphics[width=0.28\textwidth]{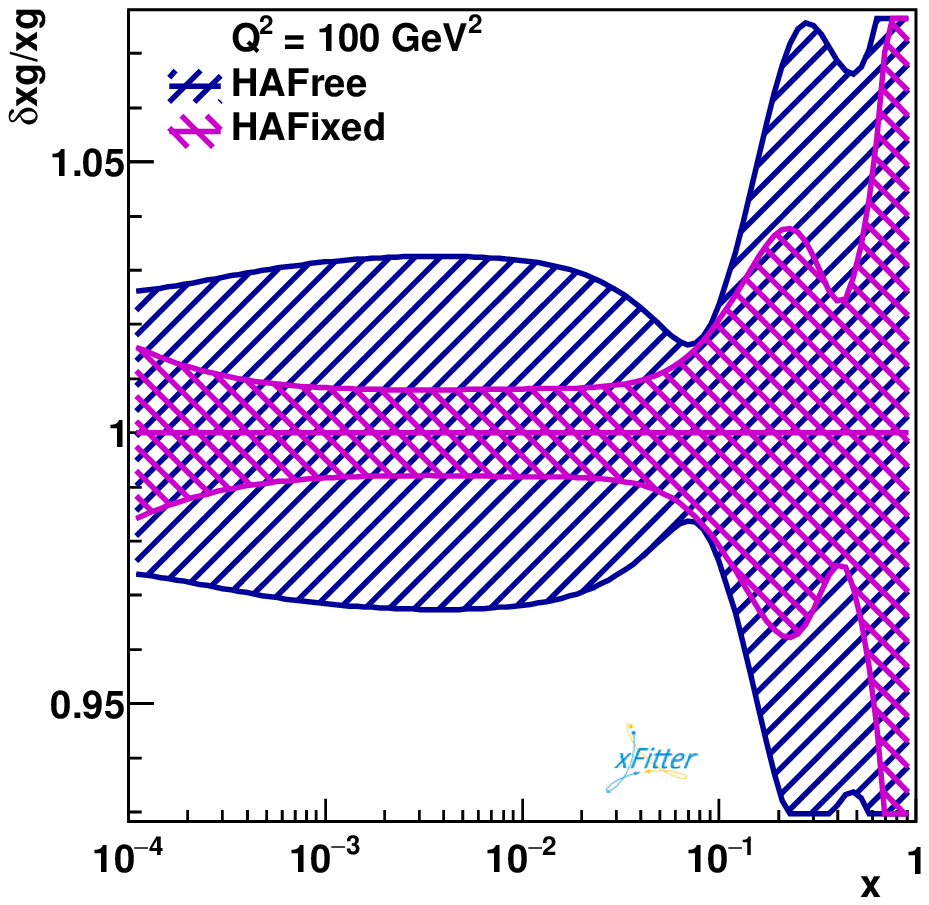}

\includegraphics[width=0.28\textwidth]{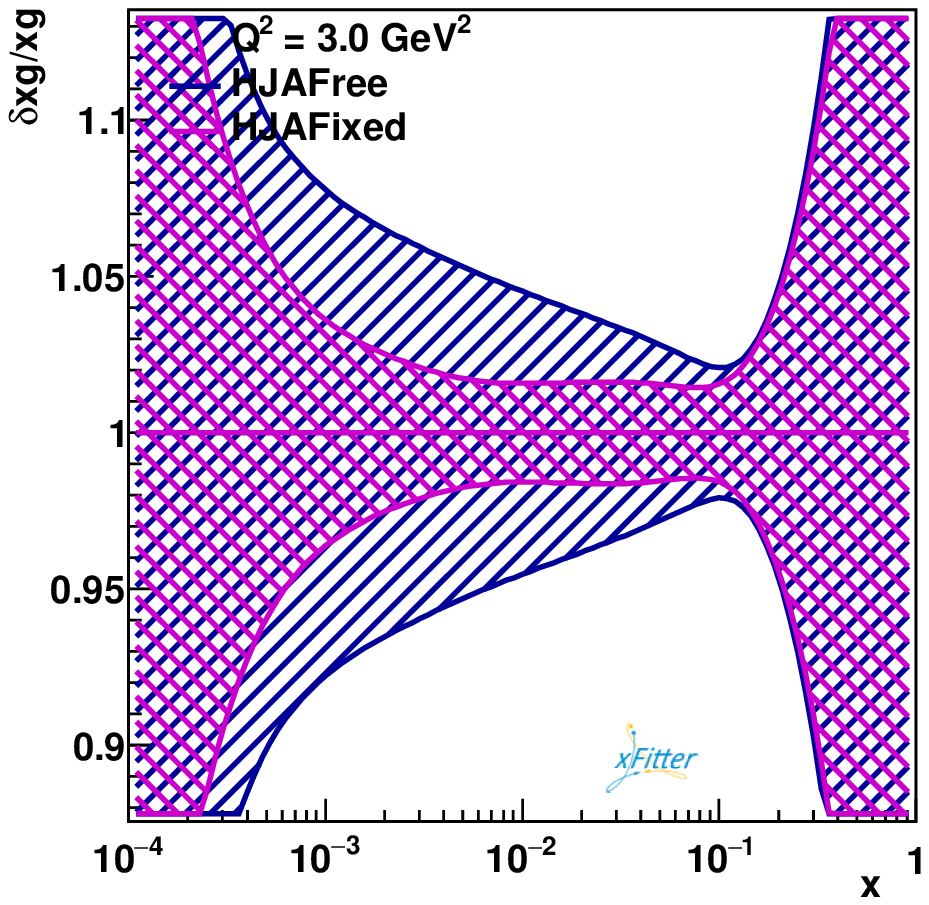}
\includegraphics[width=0.28\textwidth]{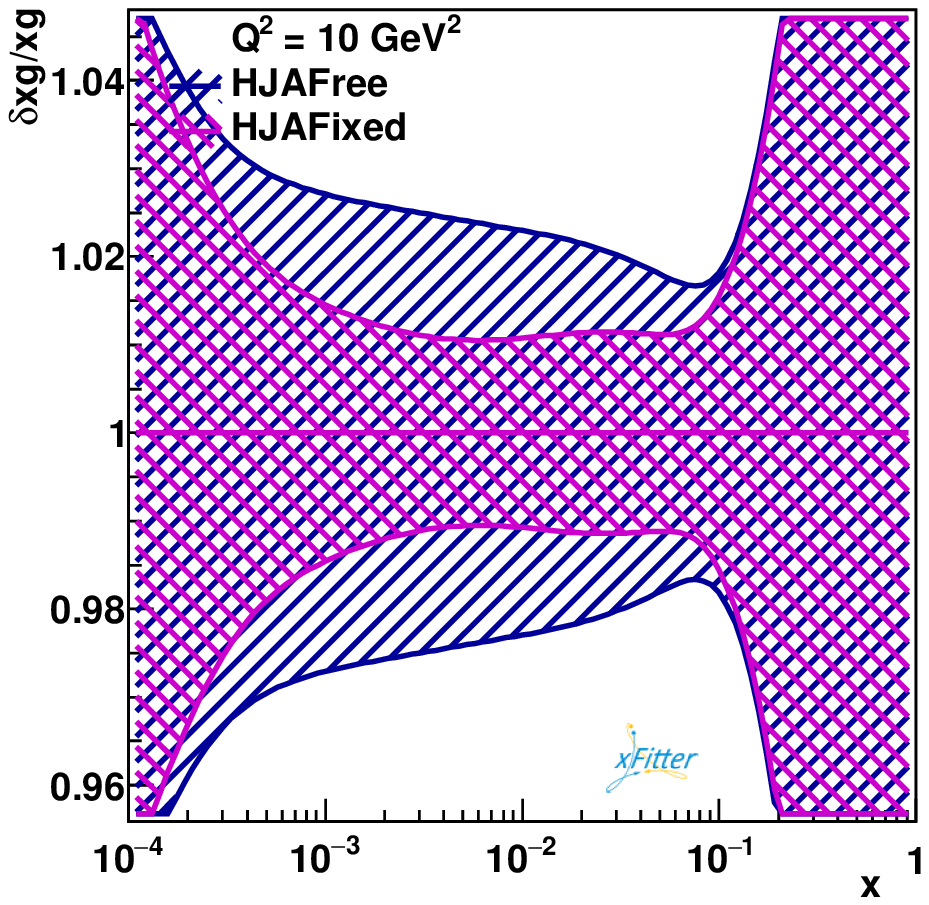}
\includegraphics[width=0.28\textwidth]{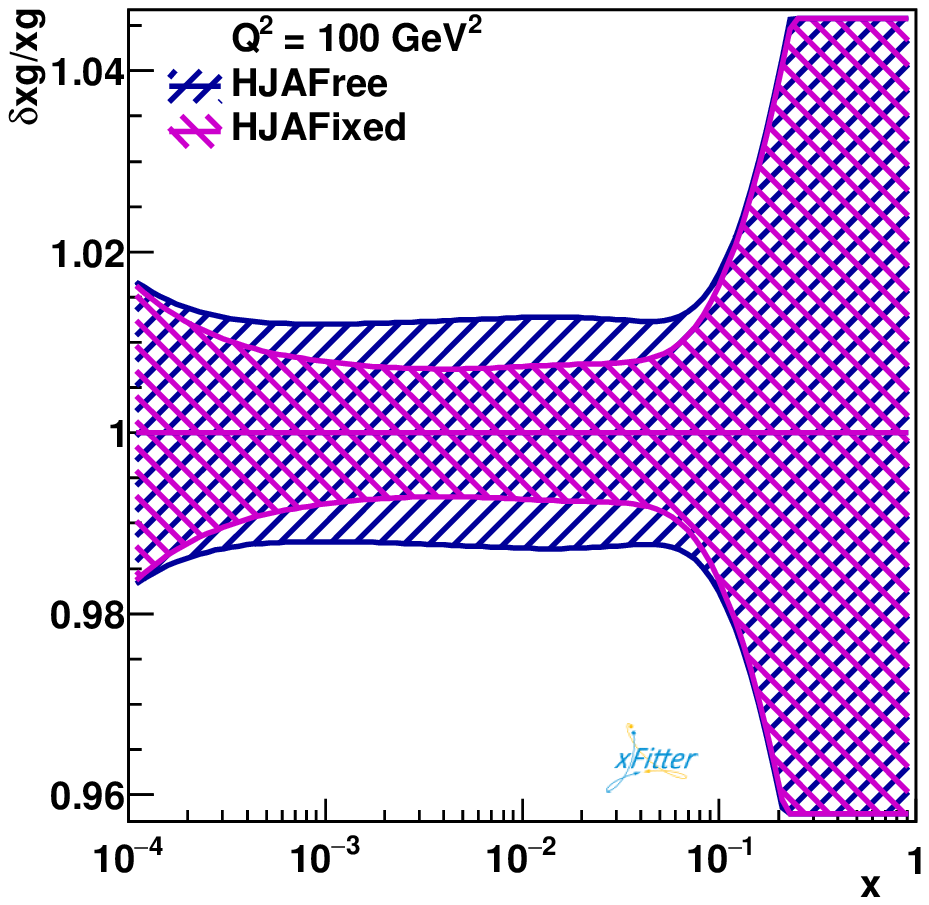}
\caption{Comparison of the pure impact of the strong coupling $\alpha_s(M^2_Z)$ on the gluon-ratio distribution for fits with fixed (purple) and free (blue) $\alpha_s(M^2_Z)$, without and with inclusion of inclusive jet production data corresponding to upper three and lower three diagrams, respectively.}
\label{fig:8}
\end{figure*}

\begin{figure*}
\includegraphics[width=0.28\textwidth]{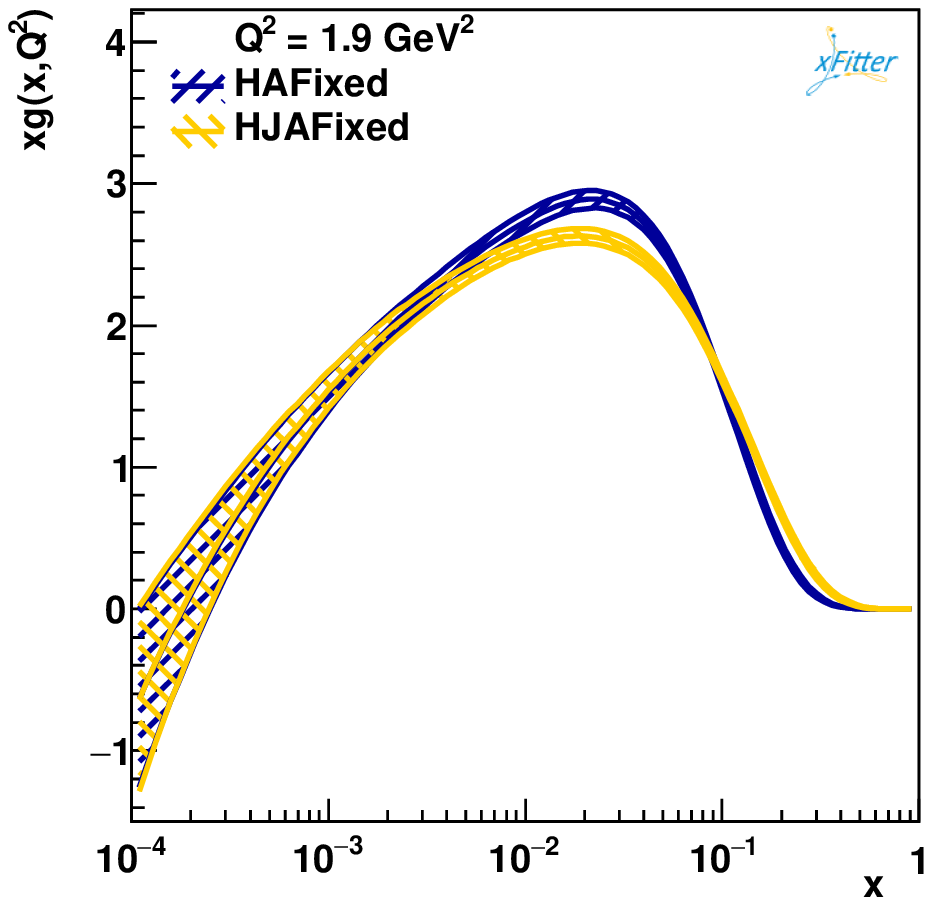}
\includegraphics[width=0.28\textwidth]{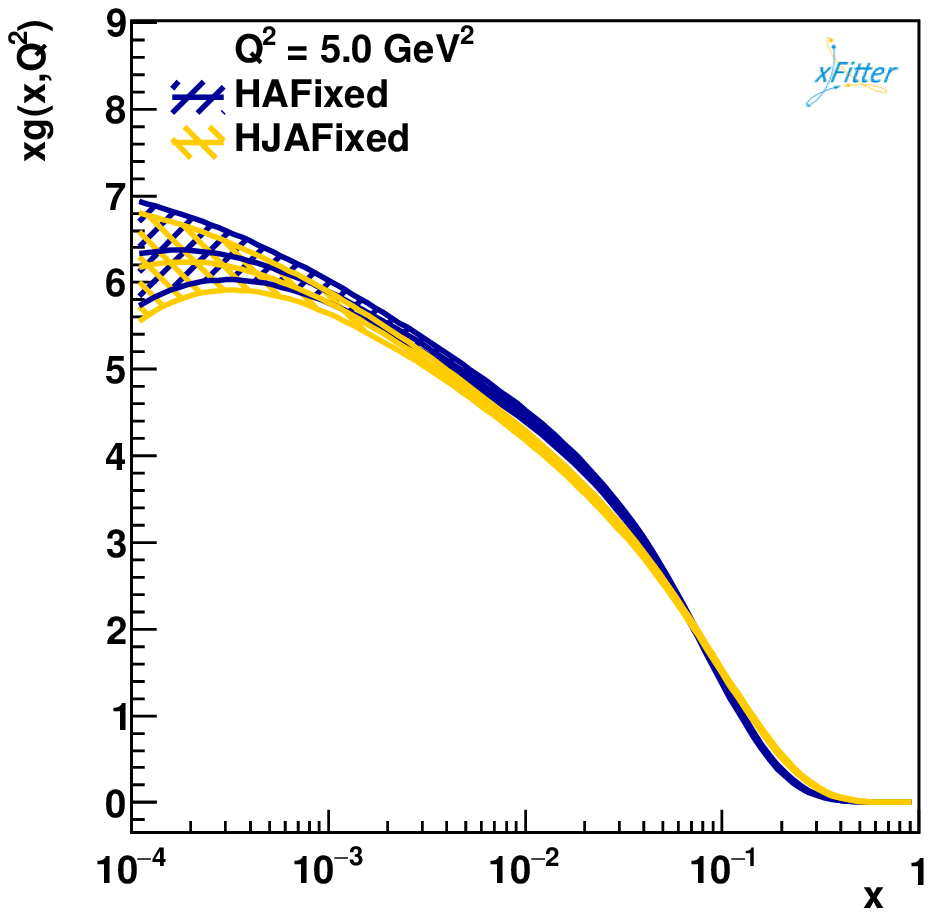}
\includegraphics[width=0.28\textwidth]{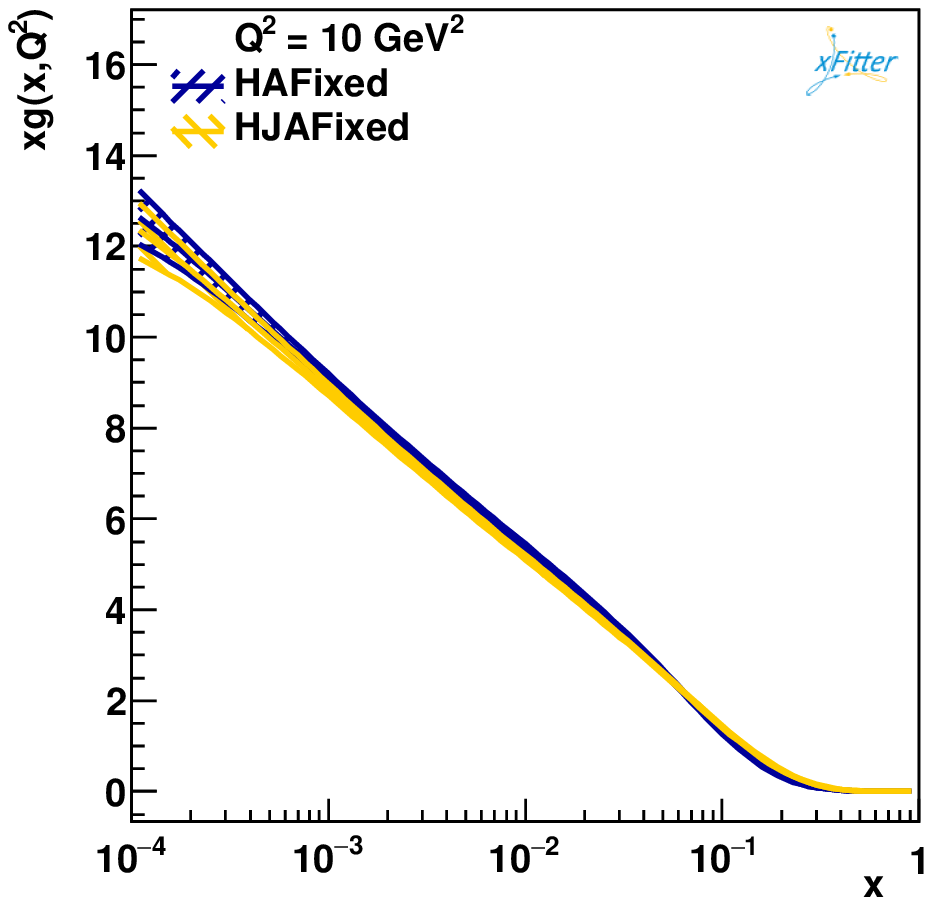}

\includegraphics[width=0.28\textwidth]{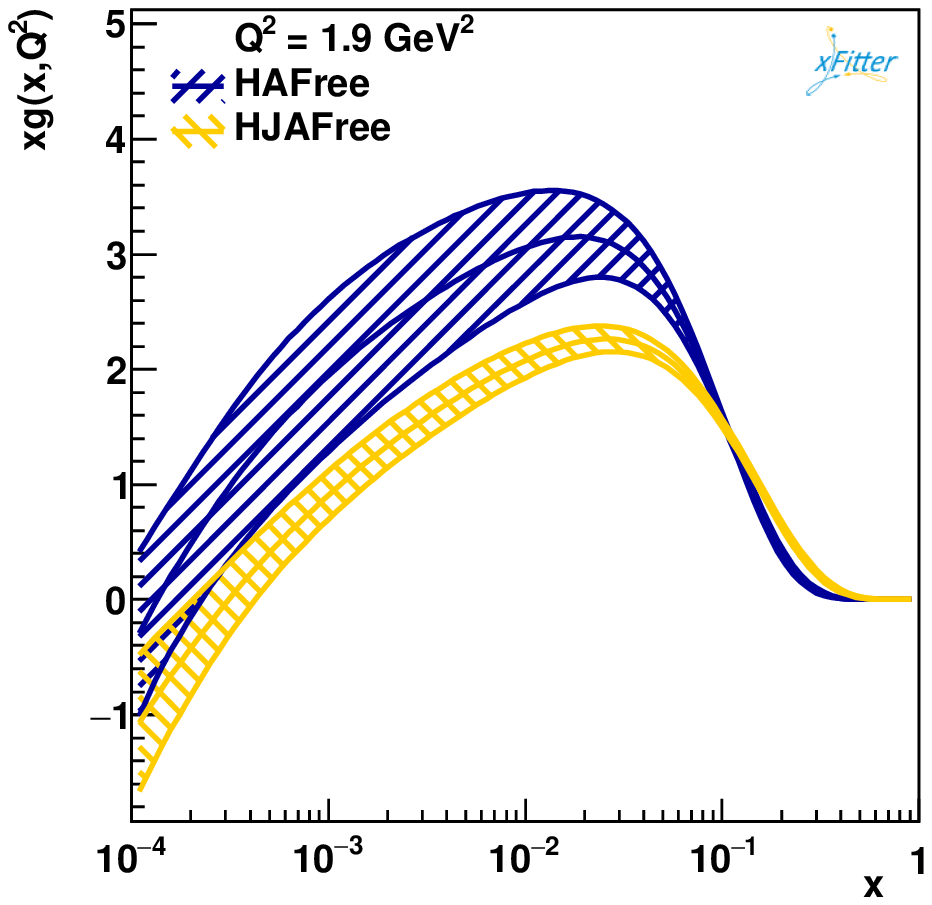}
\includegraphics[width=0.28\textwidth]{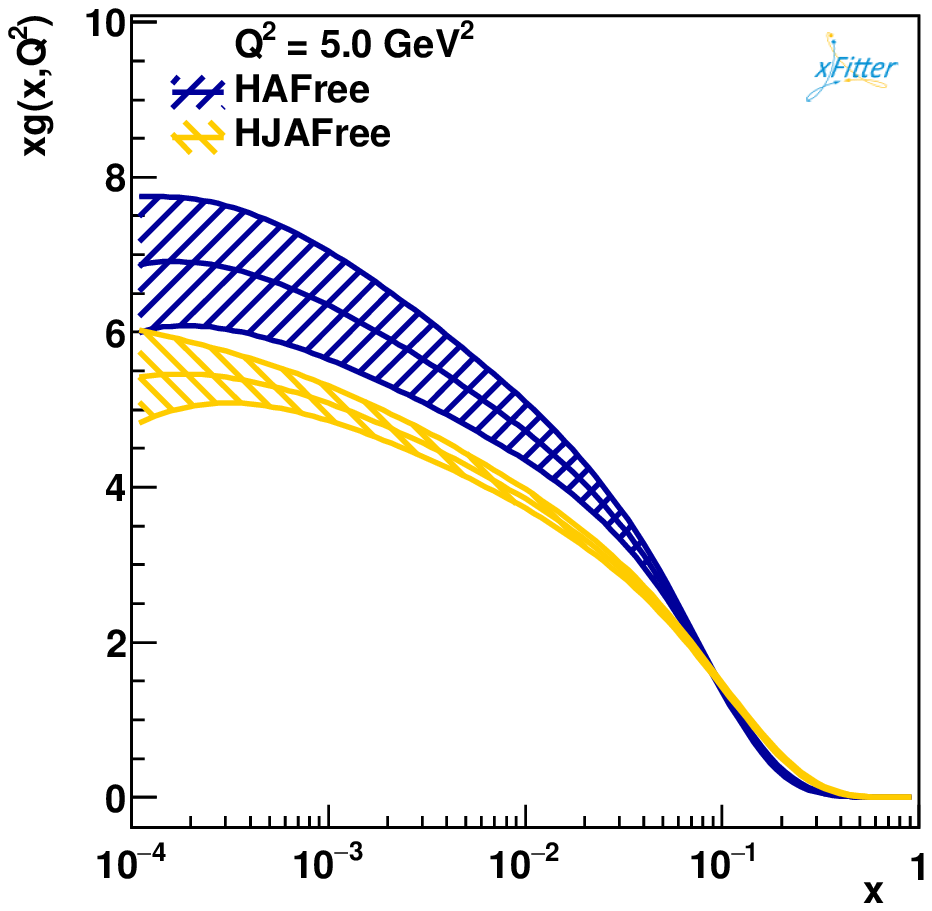}
\includegraphics[width=0.28\textwidth]{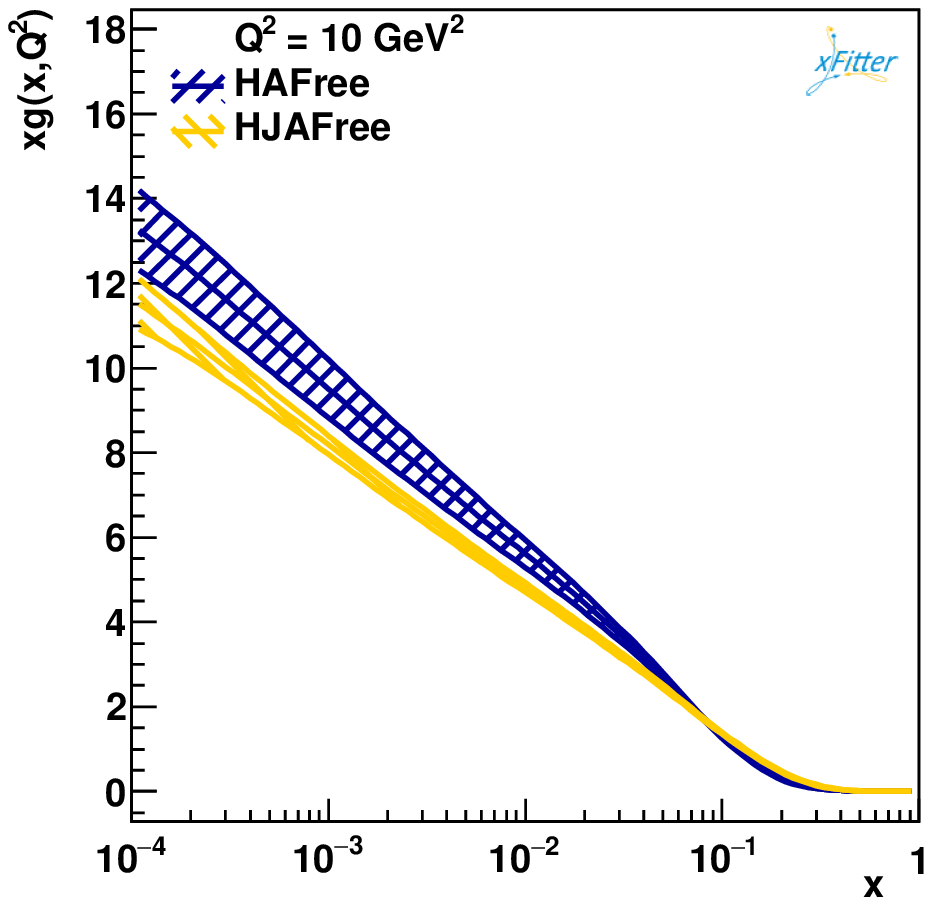}
\caption{Comparison of the pure impact of inclusion jet production data on the shape of gluon distribution for fits with fixed (upper three diagrams) and free (lower three diagrams) strong coupling $\alpha_s(M^2_Z)$.}
\label{fig:9}
\end{figure*}

\begin{figure*}
\includegraphics[width=0.28\textwidth]{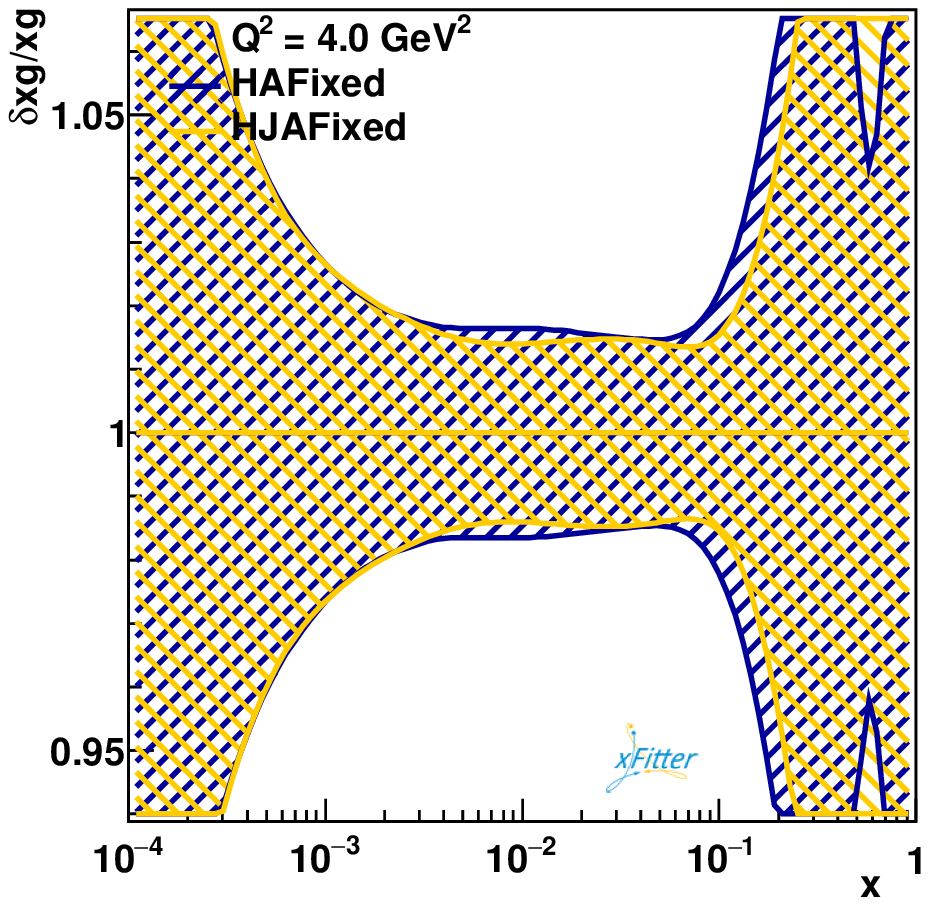}
\includegraphics[width=0.28\textwidth]{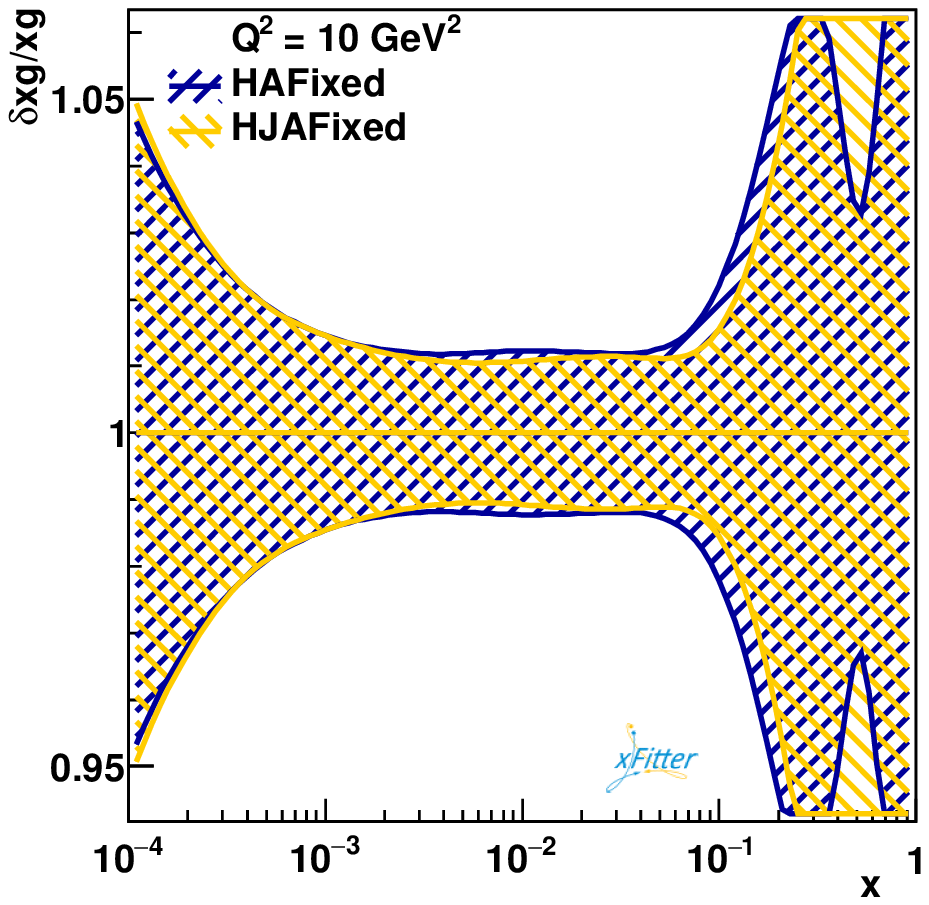}
\includegraphics[width=0.28\textwidth]{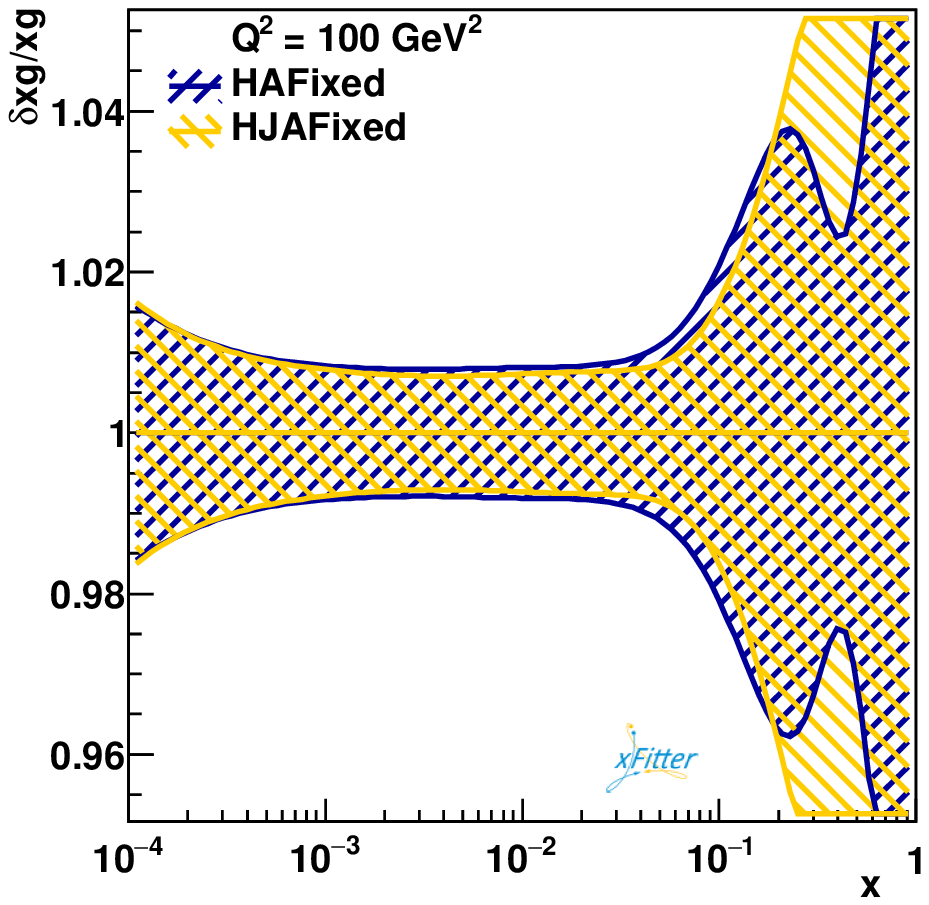}

\includegraphics[width=0.28\textwidth]{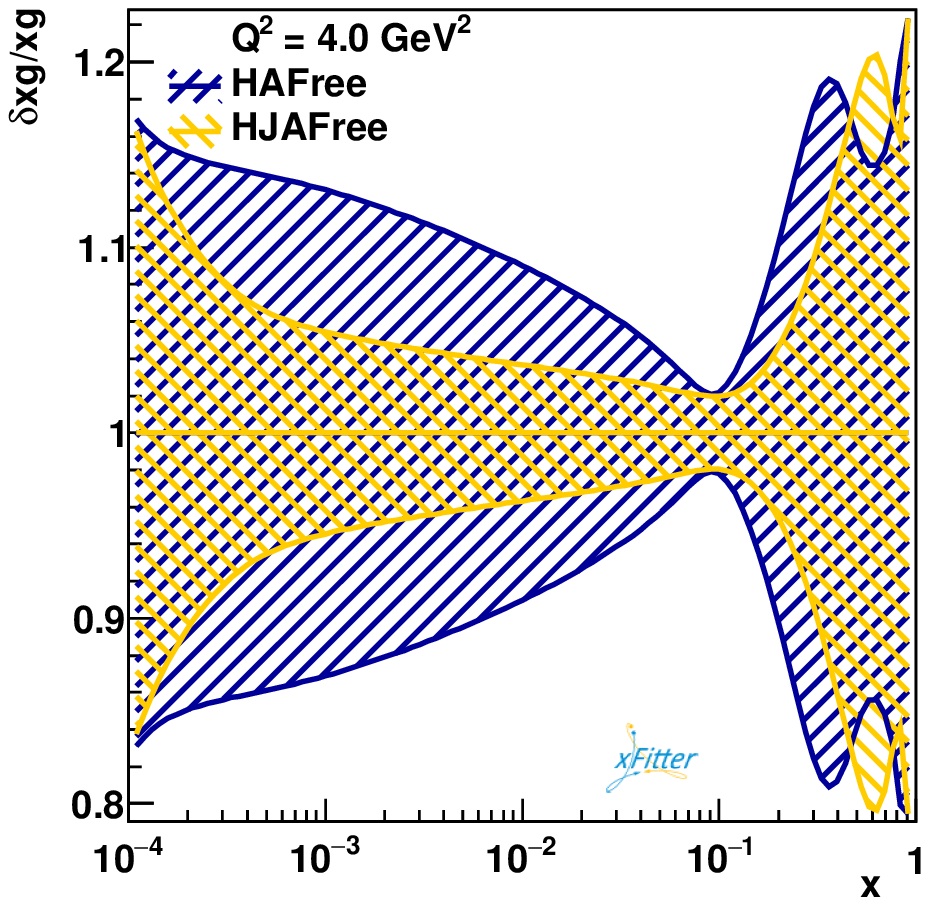}
\includegraphics[width=0.28\textwidth]{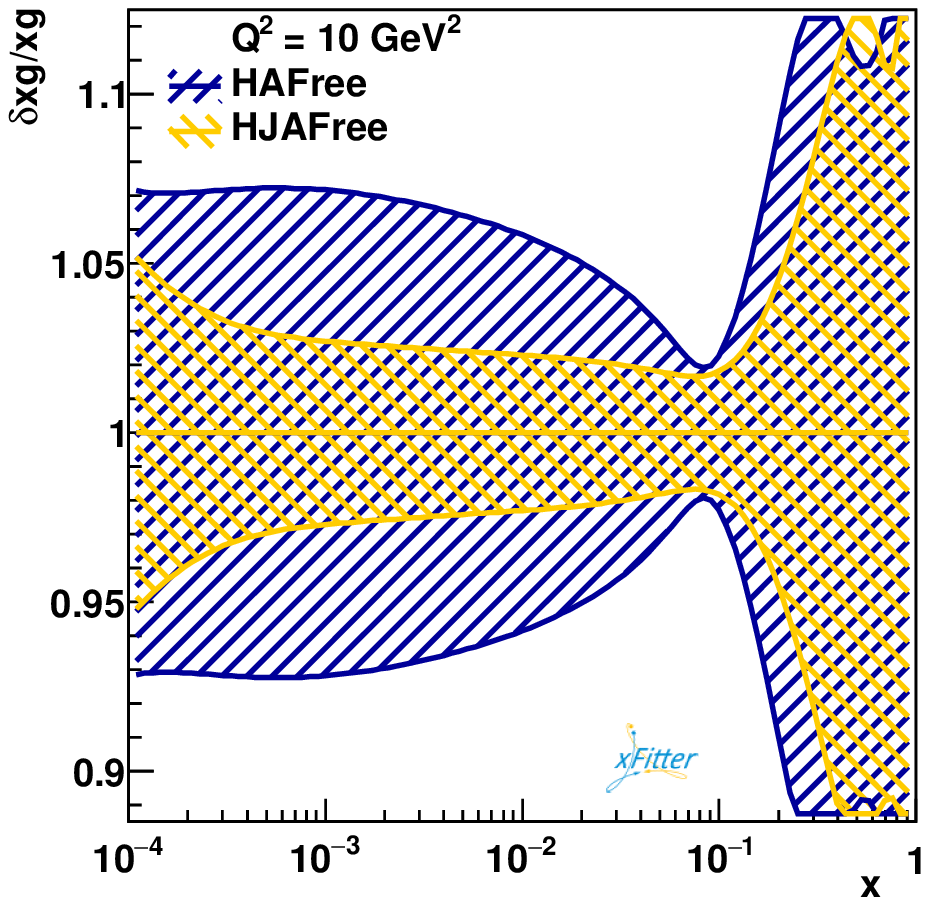}
\includegraphics[width=0.28\textwidth]{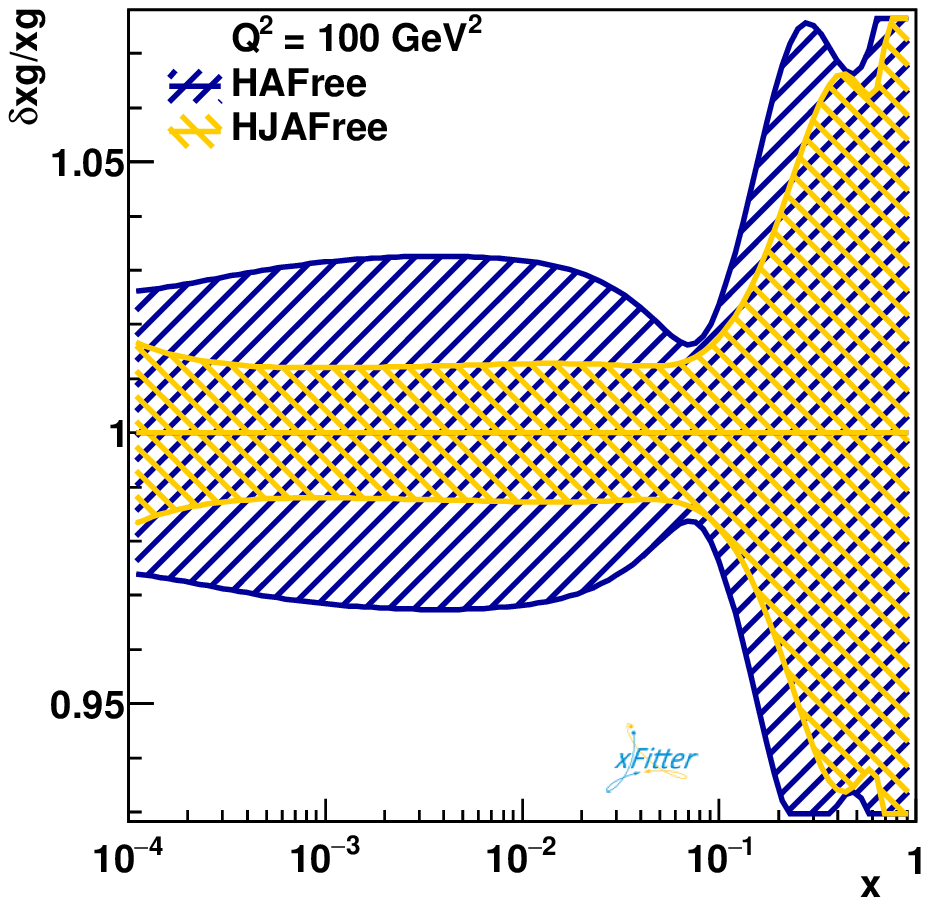}
\caption{Comparison of the pure impact of inclusion jet production data on the shape of gluon-ratio distribution for fits with fixed (upper three diagrams) and free (lower three diagrams) strong coupling $\alpha_s(M^2_Z)$.}
\label{fig:10}
\end{figure*}

\item {\bf Impact of simultaneous determination of PFDs and ${\bf \alpha_s(M^2_Z)}$:}

Figs.\ref{fig:11} and \ref{fig:12} show impact of simultaneous inclusion of jet production data and the strong coupling $\alpha_s(M^2_Z)$ on the $xu_v$, $xu_v$-ratio, $xd_v$ and $xd_v$-ratio distributions for four different HAFixed (blue), HJAFixed (orange), HAFree (green) and HJAFree (red) fits, respectively.

Impact of simultaneous inclusion of jet production data and the strong coupling $\alpha_s(M^2_Z)$ on the gluon distributions for four different HAFixed (blue), HJAFixed (orange), HAFree (green) and HJAFree (red) fits, respectively is shown in Fig.~\ref{fig:13}.

In Fig.~\ref{fig:14} we show impact of simultaneous inclusion of jet production data and the strong coupling $\alpha_s(M^2_Z)$ on the gluon-ratio distributions for four different HAFixed (blue), HJAFixed (orange), HAFree (green) and HJAFree (red) fits, respectively.

As can be seen from Fig.~\ref{fig:13}, best improvement in gluon distribution uncertainty is related to HJAFree analysis (red diagram) and this is because of strong correlation between the shape of the gluon distribution and the strong coupling $\alpha_s(M^2_Z)$, when it is considered as a free fit parameter and the central role of inclusion of jet production data in direct measurement of $\alpha_s(M^2_Z)$.

\begin{figure*}
\includegraphics[width=0.49\textwidth]{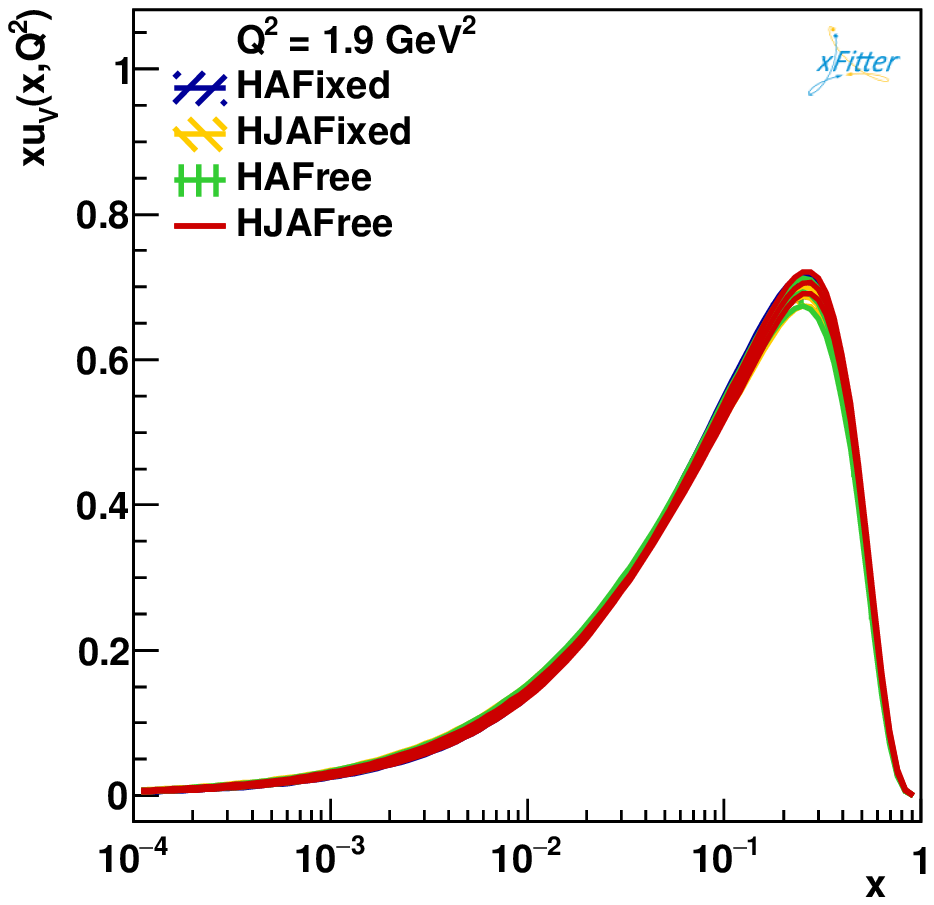}
\includegraphics[width=0.49\textwidth]{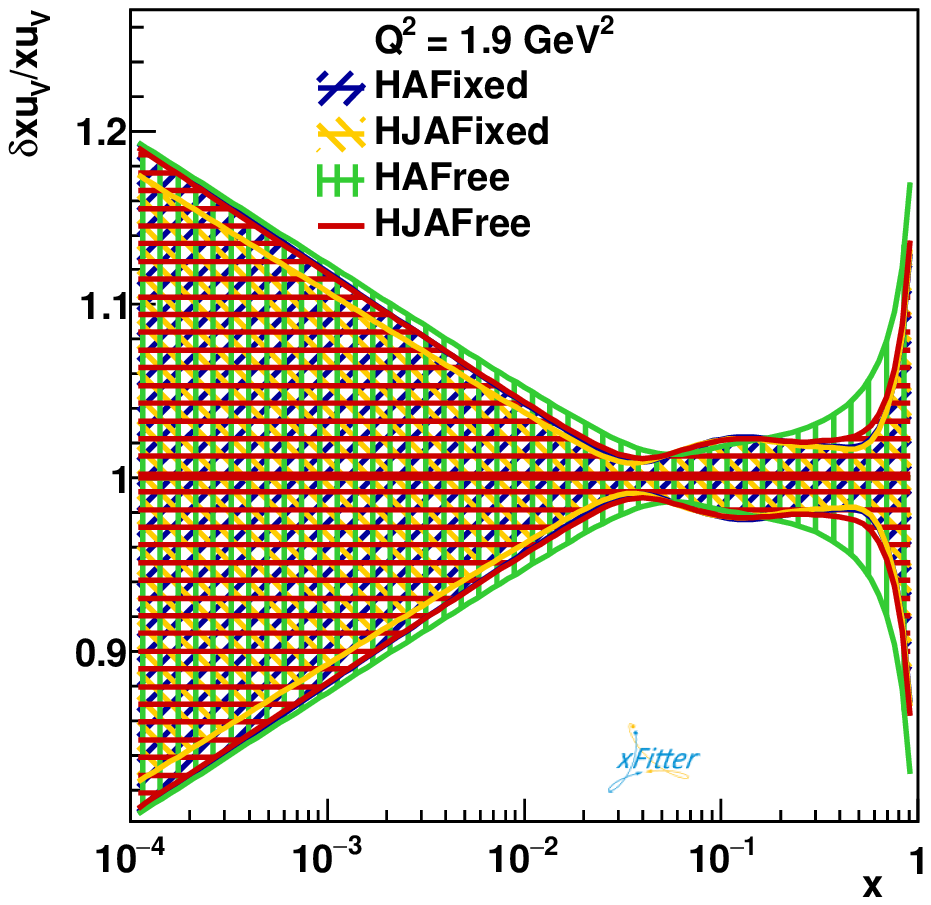}

\caption{Impact of simultaneous inclusion of jet production data and the strong coupling $\alpha_s(M^2_Z)$ on the $xu_v$ and $xu_v$-ratio distributions for four different HAFixed (blue), HJAFixed (orange), HAFree (green) and HJAFree (red) fits, respectively.}
\label{fig:11}
\end{figure*}

\begin{figure*}
\includegraphics[width=0.49\textwidth]{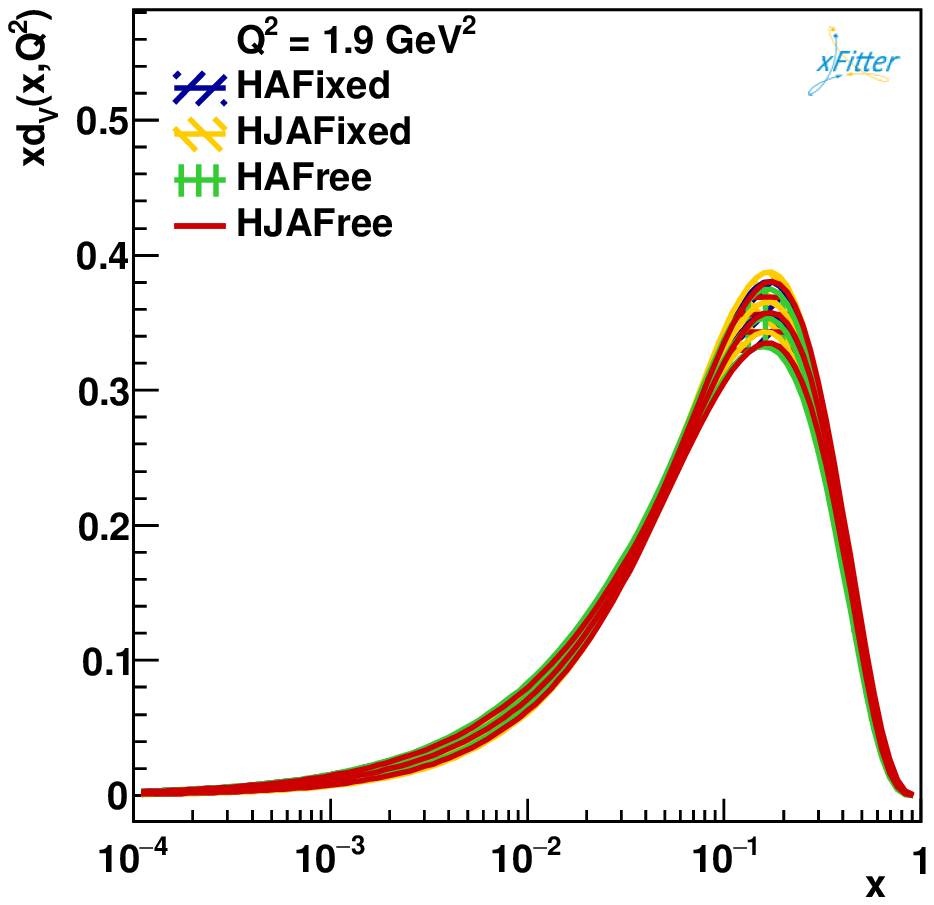}
\includegraphics[width=0.49\textwidth]{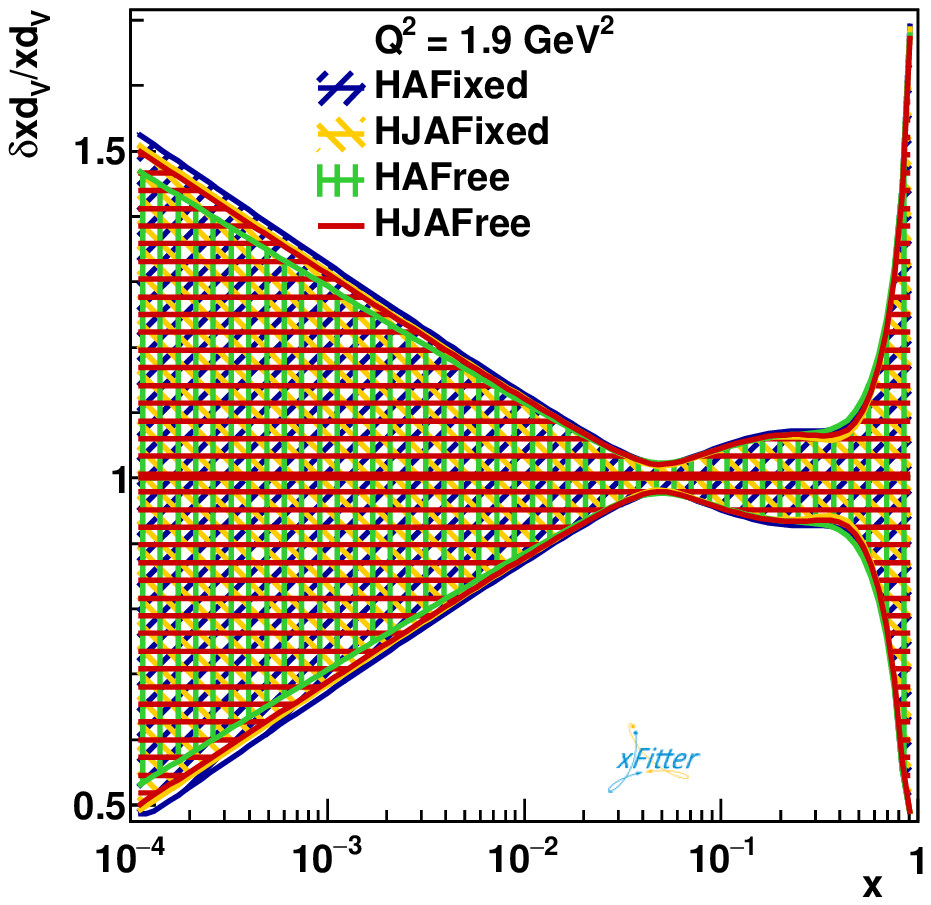}

\caption{Impact of simultaneous inclusion of jet production data and the strong coupling $\alpha_s(M^2_Z)$ on the $xd_v$ and $xd_v$-ratio distributions for four different four different HAFixed (blue), HJAFixed (orange), HAFree (green) and HJAFree (red) fits, respectively.}
\label{fig:12}
\end{figure*}

\begin{figure*}
\includegraphics[width=0.49\textwidth]{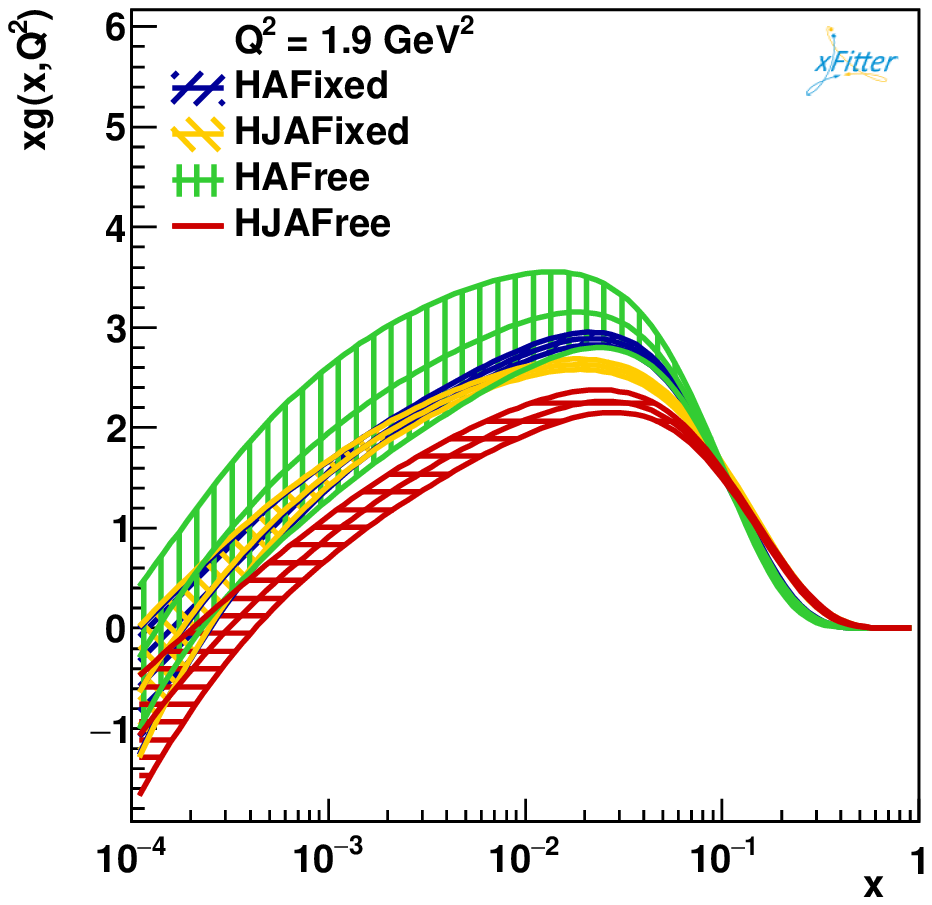}
\includegraphics[width=0.49\textwidth]{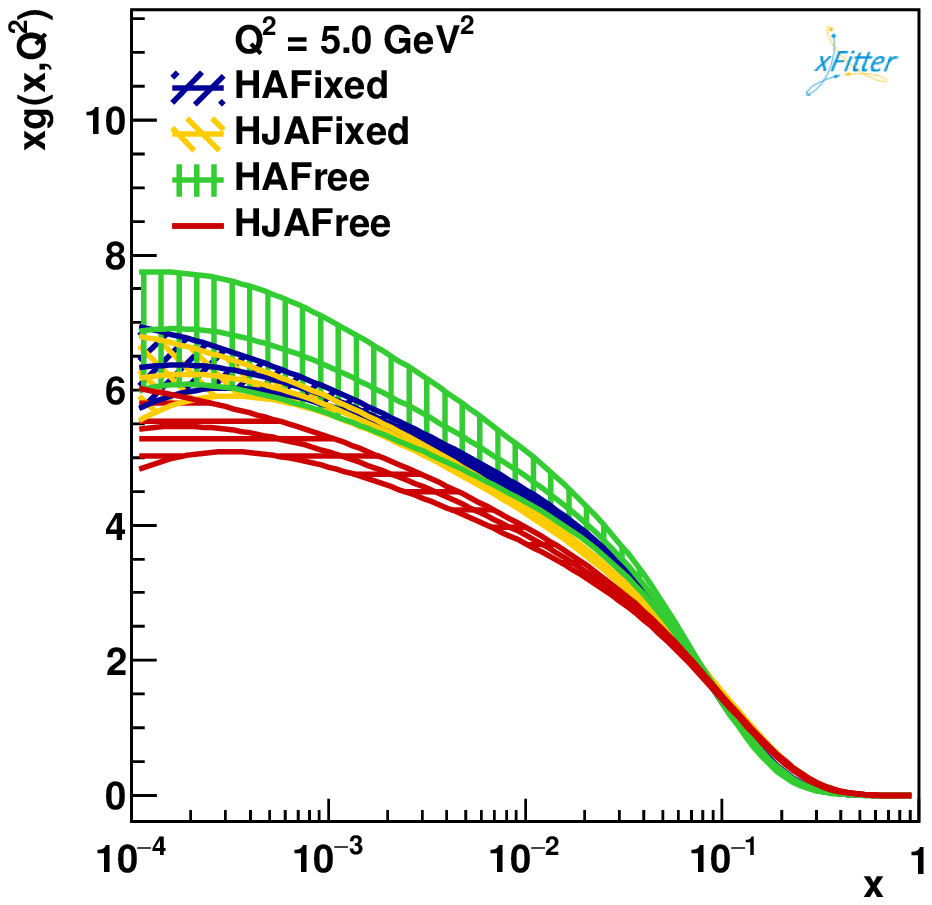}
\caption{Impact of simultaneous inclusion of jet production data and the strong coupling $\alpha_s(M^2_Z)$ on the gluon distributions for four different HAFixed (blue), HJAFixed (orange), HAFree (green) and HJAFree (red) fits, respectively.}
\label{fig:13}
\end{figure*}

\begin{figure*}
\includegraphics[width=0.49\textwidth]{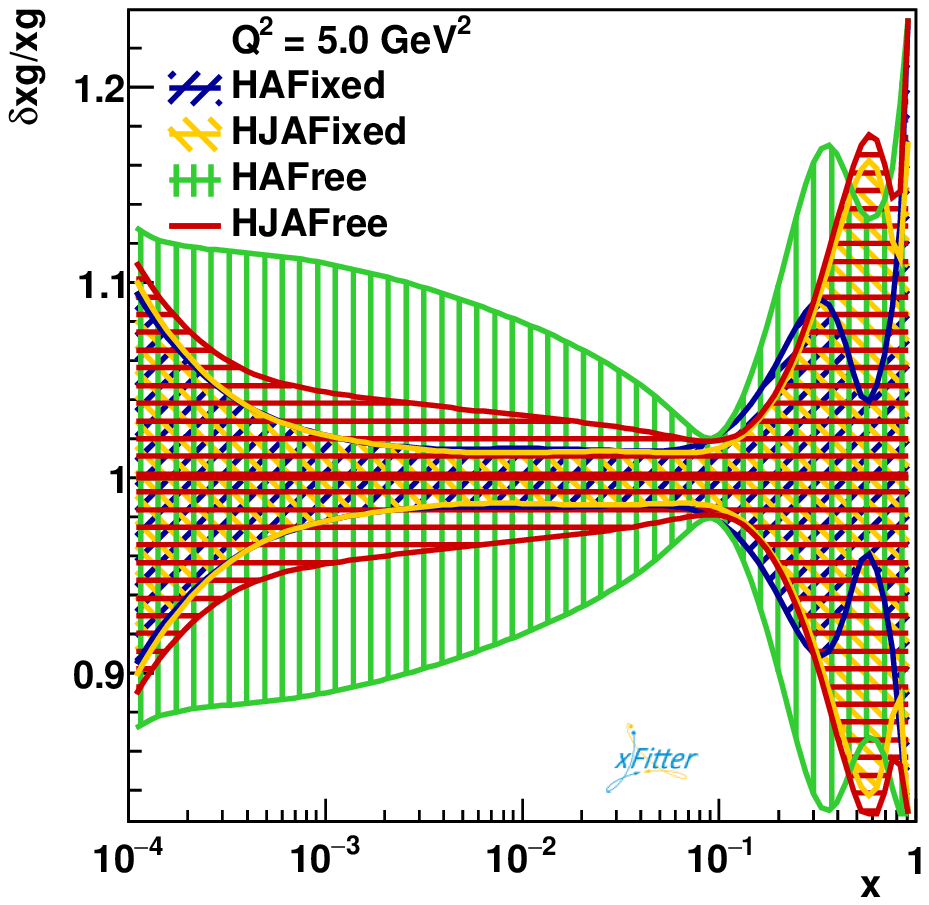}
\includegraphics[width=0.49\textwidth]{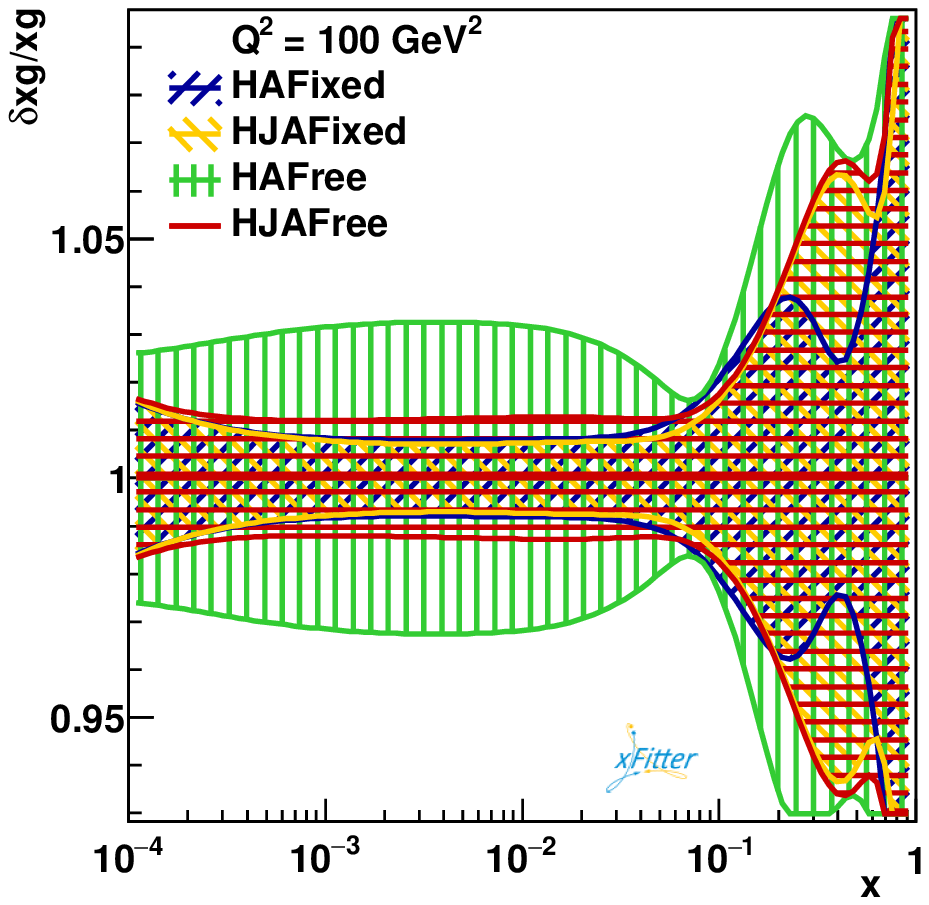}
\caption{Impact of simultaneous inclusion of jet production data and the strong coupling $\alpha_s(M^2_Z)$ on the gluon-ratio distributions for four different HAFixed (blue), HJAFixed (orange), HAFree (green) and HJAFree (red) fits, respectively.}
\label{fig:14}
\end{figure*}

\end{itemize}  

\section{\label{summary}Summary}
\begin{itemize}
\item We perform a NLO QCD analysis without and with inclusion of inclusive jet production data and the strong coupling constant $\alpha_s(M^2_Z)$ on the NC and CC deep inelastic $e^{\pm}p$ scattering cross sections theory in format of four different HAFixed, HJAFixed, HAFree and HJAFree fits.

\item We show the pure impact of inclusion of inclusive jet production data and the strong coupling constant $\alpha_s(M^2_Z)$ on the NC and CC deep inelastic $e^{\pm}p$ scattering cross sections theory is $\sim 2.8$ and $\sim 0.8$ improvement in the quality of the fit, respectively.

\item A simultaneous determination of PDFs and $\alpha_s(M^2_Z)$ with inclusion of inclusive jet production data on the HERA I and II combined data leads to $\sim 3.6$ improvement in the quality of the fit and gives the numerical values of strong coupling as: $\alpha_s^{{\rm NLO}}(M^2_Z) = 0.1160 \pm 0.0049$ and $\alpha_s^{{\rm NLO}}(M^2_Z) = 0.12041 \pm 0.00086 $ corresponding to HAFree and HJAFree analysis.

\item Because of strong correlation between the shape of the gluon distribution and the strong coupling $\alpha_s(M^2_Z)$, when it is considered as a free fit parameter and the central role of inclusion of jet production data in direct measurement of $\alpha_s(M^2_Z)$, the best improvement in gluon distribution uncertainty is related to HJAFree analysis.       
\end{itemize}



\end{document}